\documentclass[letterpaper]{article} % DO NOT CHANGE THIS
\usepackage{aaai2026}  % DO NOT CHANGE THIS
\usepackage{times}  % DO NOT CHANGE THIS
\usepackage{helvet}  % DO NOT CHANGE THIS
\usepackage{courier}  % DO NOT CHANGE THIS
\usepackage[hyphens]{url}  % DO NOT CHANGE THIS
\usepackage{graphicx} % DO NOT CHANGE THIS
\usepackage{natbib}  % DO NOT CHANGE THIS AND DO NOT ADD ANY OPTIONS TO IT
\usepackage{caption} % DO NOT CHANGE THIS AND DO NOT ADD ANY OPTIONS TO IT
\usepackage{algorithm}
\usepackage{algorithmic}

\usepackage{newfloat}
\usepackage{listings}
\DeclareCaptionStyle{ruled}{labelfont=normalfont,labelsep=colon,strut=off} % DO NOT CHANGE THIS
\floatstyle{ruled}
\newfloat{listing}{tb}{lst}{}
\floatname{listing}{Listing}
\usepackage{booktabs}
\usepackage{adjustbox}
\usepackage{multirow}
\usepackage{amssymb}
\usepackage{amsmath}
\usepackage{subcaption}
\usepackage{comment}
\usepackage[flushleft]{threeparttable}
\usepackage{array}
\usepackage{longtable}
\usepackage{tabularx}
\usepackage{ragged2e}

\usepackage{etoolbox}
\usepackage{color}
\newtoggle{comments}
\toggletrue{comments}
\iftoggle{comments}{
    \newcommand{\noteeh}[1]
  {{\color{magenta}[{\bf Emma:} #1]}}
  \newcommand{\todo}[1]
  {{\color{red}[{\bf Todo:} #1]}} 
  \newcommand{\cut}[1]
  {{\color{red}\sout{#1}}}  
  
}{
  \newcommand{\noteeh}[1]{}
  \newcommand{\todo}[1]{}
  \newcommand{\cut}[1]{}
  
}

\newcolumntype{M}[1]{>{\raggedright\arraybackslash}m{#1}}

\urldef\ttaxurl\url{https://www.cookcountyclerkil.gov/sites/default/files/pdfs/Transfer%20Tax%20By%20Municipality%2C%20January%202025a.pdf}

\setcounter{secnumdepth}{2} %May be changed to 1 or 2 if section numbers are desired.

\title{Hidden Errors in Big Data: The Case of Property Records}
\author {
    Evelyn Smith\textsuperscript{\rm 1},
    Emma Harvey\textsuperscript{\rm 2},
    Jacob S. Goldin\textsuperscript{\rm 3},
    Daniel E. Ho\textsuperscript{\rm 4}
}
\affiliations {
    \textsuperscript{\rm 1}American Bar Foundation\\
    \textsuperscript{\rm 2}Cornell Tech\\
    \textsuperscript{\rm 3}University of Chicago, co-equal supervising author\\
    \textsuperscript{\rm 4}Stanford University, co-equal supervising author\\
    esmith@abfn.org, evh29@cornell.edu, jsgoldin@uchicago.edu, deho@law.stanford.edu
}

\begin{document}

\maketitle

\begin{abstract}
Big data are the foundation for an increasing share of academic research and AI models deployed in both the public and private sectors, prompting substantial growth over time in reliance on brokered datasets. Brokered property records, which are ubiquitous in studies of gentrification, inequality, and the property tax in the U.S. and serve as inputs to property valuation models, are one notable example. In this paper, we audit two prominent brokered property datasets, finding errors in these data which bias key measures of economic inequality. First, we document that for 1-2\% of matched sales in Cook County, IL, from 2018-2021, broker-provided sale prices differ from ground truth sale prices by more than 5\%. Moreover, missing data and conceptual differences in the reporting of deed and property characteristics lead to coverage errors ranging from 12 to 15\% of transactions. Second, we show that misreporting is highly consistent between brokers: more often than not, brokers make identical reporting errors for the same transactions. Third, to illustrate the significance of these errors, we measure their impact on estimates of property tax regressivity, finding that they drive significant wedges between estimates depending on the data source. 
These findings generalize to two other large counties in the U.S., and highlight the crucial importance of open administrative data and transparency from brokers regarding data provenance and lineage.\looseness=-1

%We have confirmed these discrepancies with the data broker, and until corrections can be issued, we offer heuristics to assess robustness across published estimates relying on these data. 
\end{abstract}


\section{Introduction} \label{intro}

Big data now comprise a substantial share of the evidence base for disciplines ranging from medicine and genomics to climatology and the social sciences, while also serving as an important empirical basis for policymaking, law enforcement, and the machine learning systems that increasingly mediate government decisionmaking \citep{martin2014big, o2013big, mangal2020big, einav2014economics, burrows2014after, lane2016big, coulthart2022putting, hu2015big, odni_cai_framework_2024, wade2023clocks}. While large, complex datasets have advanced our understanding of important topics, their scale and volume complicate their growing role in scientific research and public policy, making it difficult for end users to validate these data for accuracy and representativeness. These challenges reinforce a broader concern that big data yield estimates which are simultaneously precise and misleading\textemdash a tension which \citet*{meng2018statistical} terms the ``Big Data Paradox.'' As governments  adopt big data and machine learning tools for tasks ranging from law enforcement to property assessment, data quality becomes an essential determinant of system outcomes and thus an important policy concern; indeed, as \citet{sambasivan2021everyone} note, data errors cascade through AI systems, making these systems harder to maintain, harming the individuals these systems serve, and eroding institutional trust.\looseness=-1

Accounting for big data's provenance and lineage is a particular challenge \citep{wang2015big, chacko2017big, appelbaum2016securing}. This is especially true for records sourced from data brokers, who make crucial and often opaque decisions regarding how to source, clean, impute, and formalize their data. These decisions can introduce consequential errors if brokers disregard subtleties or key institutional details in their data sources \citep{zook2023crisis, donaldson2024transparency}. Moreover, data brokers frequently sell data to one another, creating opportunities for errors to propagate across networks of brokers \citep{FTCdatabrokers2014}. These limitations are particularly alarming in light of the growing literature on data documentation for machine learning and pipeline-aware bias mitigation, which argue that downstream model behavior cannot be properly accounted for without careful scrutiny of upstream data practices \citep{gebru2021datasheets, pushkarna2022data, black2023toward, suresh_framework_2021}.\looseness=-1

In this paper, we audit the accuracy and coverage of two prominent property data brokers whose datasets are widely used in government and academic research: Cotality and ATTOM. Both brokers advertise comprehensive or near-comprehensive coverage of properties, mortgages, sales, and foreclosures across the U.S., and their data are ubiquitous in research on housing markets, as shown in Appendix Table \ref{tab:cotality_papers}.\footnote{ATTOM claims their data cover more than 158 million parcels; Cotality advertises 99.9\% market coverage.} Cotality and ATTOM also play a growing role in housing policy and property assessment. ATTOM, for example, has contracted with the Department of Housing and Urban Development to provide foreclosure and sales data to inform agency policy, while both Cotality and ATTOM provide data and software, including ML-based valuation models, that counties can use to perform assessments \citep{sanmateo_property_appraisal_2024, cotality_marshall_swift,hud_noi_attom_2024}. These brokers are central to the administration of private-sector systems as well, such as mortgage financing and private appraisals. Cotality and ATTOM source their data in part from small units of local government, including county recorders and assessors, who number in the thousands and have diverse standards and practices for data reporting. To produce usable property and transaction data at the scale that they advertise, brokers must encode key concepts such as foreclosures and arms-length sales across tens of millions of inconsistent records\textemdash a laborious undertaking for any institution. However, as brokers generally do not publicize their data lineage, end users cannot account for the accuracy of these encodings or for errors that occur as concepts are harmonized across jurisdictions, even though these errors may have significant consequences for model validity \cite{jennifer2021garbage}.\looseness=-1

Our audit identifies substantial coverage and imputation errors in Cotality and ATTOM relative to administrative data for three jurisdictions and a sample period covering 2018-2021. After matching transactions of single-family homes between brokered and administrative data, we find that broker-reported sale prices substantially differ from county-reported prices for 1-2\% of matched sales. These differences are on the order of hundreds of thousands of dollars per transaction, and are likely attributable to broker error in the process of imputing sale prices from transfer tax payments. Moreover, we identify coverage error ranging from 12-15\% of ground truth transactions, largely owing to missing data and broker misreporting of filters used to subset to arms-length transactions of single-family homes. We find these errors significantly affect broker-derived measures of property tax assessment regressivity, a prominent metric in ongoing academic and policy debates.\looseness=-1

We provide an overview of the data in \S\ref{data}. In \S\ref{findings}, we identify the sources and extent of error separately for ATTOM and Cotality. \S\ref{error-concordance} compares errors across brokers, examining to what extent these errors overlap, and \S\ref{implications} examines the implications of these errors for property tax assessment regressivity estimates.\looseness=-1

\section{Background and Related Work}
Our work contributes to a landscape of responsible AI research that considers ``data-centric factors''~\cite{li_data_2022} in AI development and evaluation, with a particular focus on how transparent data practices can promote accountability for data owners and data users~\cite{Chappidi_Cobbe_Norval_Mazumder_Singh_2025}. In line with prior work in this space, we interrogate large-scale, widely-used datasets~\cite{Agnew_Barnett_Chu_Hong_Feffer_Netzorg_Jiang_Awumey_Das_2025} in order to make the practices underlying data development clear~\cite{Rothschild_Wang_JayakumarVilvanathan_Wilcox_DiSalvo_DiSalvo_2024, zajkac_ground_2023}.\looseness=-1

More specifically, our work contributes to a prior literature examining the impact of sampling, coverage, and imputation errors on estimates of significant social importance derived from big data, including vaccine uptake \citep{bradley2021unrepresentative}, influenza incidence \citep{lazer2014parable}, electoral polling \citep{meng2018statistical}, and the relationship between cardiovascular disease and COVID-19 mortality \citep{mehra2020cardiovascular}. We also build on previous research auditing brokered datasets which has identified considerable discrepancies across brokers in the quality and coverage of their data, particularly in the context of mobility data \citep{hsu2024human, noi2022assessing}. The most closely related body of research this paper examines the challenges of standardizing large-scale real estate microdata \citep{nolte2024data, brummet_analysis_2016, seeskin_evaluating_2018}. Consistent with the findings we present here, these prior papers identify substantial discrepancies between broker-provided and administrative records for features including home ownership data, mortgage data, and property tax liabilities, the scope and magnitude of which vary across geographies, and which are attributable to conceptual differences across jurisdictions and between brokers and governments in data collection and reporting. Our paper is the first to compare distinct brokered property databases to one another, identifying substantial similarities in errors made and measuring the impact of these errors on estimates of key statistics.\looseness=-1
\section{Data} \label{data}

Our data comprise the universe of arms-length, single-family home transactions in Cook County, IL listed in Cotality, ATTOM, and Cook County's open data repository for a period covering 2018-2021.\footnote{Our analysis is focused on Cook County because most jurisdictions do not provide property data that is publicly available for download at scale. We use comparable data from New York City, NY, and Philadelphia, PA in robustness checks in Appendix \ref{appendix:nyc}.} In each case, we exclude transactions with sale prices less than \$10K, as well as multi-parcel sales and properties which transacted more than once in a given year, which may include duplicate sales and recording errors.\footnote{A ``multi-parcel sale" is a sale involving more than one property. These sales are excluded because the recorded price reflects the value of multiple properties.} We also exclude transactions that are missing address, sale date, or sale amount information. Our filtered Cook County data contain 152,060 transactions from ATTOM, 147,731 transactions from Cotality, and 153,044 transactions from the ground-truth data for the whole of our sample period. Full details regarding the filters applied to the Cook County open data are provided in Appendix \ref{app:data_proc}.\looseness=-1

\paragraph{Treating county data as ground truth.}

Throughout, we refer to county data as ``ground truth," as these data are derived directly from county administrative records and form the basis of brokered datasets. Even so, administrative records may contain errors, and discrepancies between brokered and administrative data may arise if brokers correct these errors while processing and standardizing records across jurisdictions. To support our interpretation of county data as ground truth, we manually verify broker and county prices against deed records published by the Cook County Clerk for a random 2\% sample of transactions where the two sources disagree. For discrepancies with Cotality, the deed price matches Cook County's record 89.2\% of the time, yielding an overall County accuracy of 99.8\%.\footnote{This represents 33 of 37 cases. Of the remaining four cases, three match Cotality's data, and one involves two deeds issued several days apart\textemdash one consistent with the county's price and one with Cotality's.} Similarly, for discrepancies with ATTOM, the deed price matches Cook County's record 97.2\% of the time, yielding an overall County accuracy of 99.9\%.\footnote{This represents 36 out of 37 cases. In the remaining case, the deed price matches ATTOM's data, while Cook County's price corresponds to the mortgage amount. In two instances, disagreement between ATTOM and county data arose due to corrections made after deeds were published. In these cases, ATTOM's price reflected the original price, while Cook County's data matched the corrected price, meaning that ATTOM data was more stale than County data and was likely not updated after initial ingestion.} Altogether, for the vast majority of disagreements, the county price aligns with the most recent underlying deed record, confirming that discrepancies between sources are predominantly attributable to broker processing rather than county-level errors. While a small number of cases do favor brokered data\textemdash suggesting that brokers may occasionally correct genuine administrative mistakes\textemdash such instances are rare enough that they are unlikely to materially affect our analysis.\looseness=-1

\paragraph{Data appear comparable in aggregate.}

\begin{figure*}[h!]
    \centering

    \includegraphics[width=\linewidth]{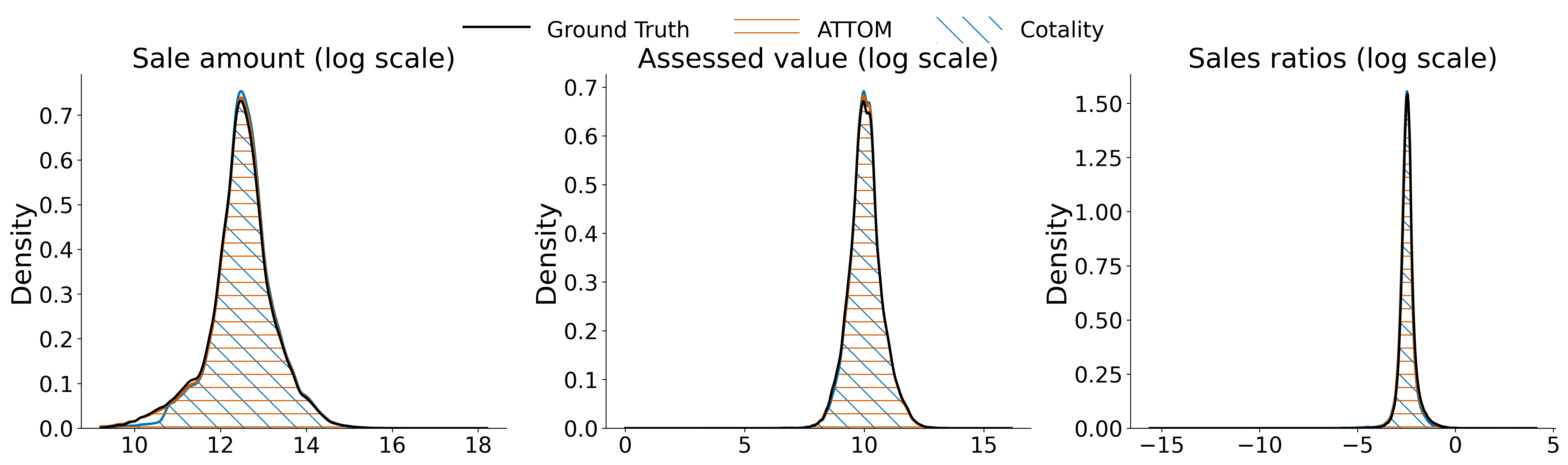}
    \caption{\textbf{Distribution of Sale Price, Assessed Value, and Assessment Ratios by Data Source, 2018-2021.} The figure shows the distributions of sale price, assessed value, and assessment ratios by data source regressivity measures by year and data source for a pooled sample of single-family home sales in Cook County covering 2018-2021. ``Ground truth" values are derived from data published by the Cook County Assessor's office.}
    \label{fig:density_plots}
\end{figure*}

Table \ref{tab:summ_stats} and Figure \ref{fig:density_plots} compare the distributions of price, assessed value, and assessment ratios across data sources for Cook County. In aggregate, the three data sources appear comparable: there are no visible differences between the densities, apart from a slight leftward skew in Cotality's sale price distribution and a corresponding rightward skew in its distribution of assessment ratios. It is not apparent simply by comparing these densities that the two brokered datasets contain discrepancies relative to ground-truth records, or that these discrepancies would have statistically significant effects on estimates derived from these data. However, as shown in subsequent sections, these summary distributions belie significant errors at the transaction level.\looseness=-1

\begin{table*}[]
\small
\centering
\setlength\tabcolsep{0.04\linewidth}

\begin{tabular}{llll}
\toprule
 & \textbf{Ground Truth} & \textbf{ATTOM} & \textbf{Cotality} \\
\midrule
N & 153,044 & 152,060 & 147,731 \\
Mean Sale Price & \$350,405 & \$356,661 & \$367,500 \\
Sale Price P1 & \$27,342 & \$25,000 & \$45,000 \\
Sale Price P25 & \$181,000 & \$186,000 & \$195,000 \\
Sale Price P50 & \$270,000 & \$275,000 & \$280,000 \\
Sale Price P75 & \$400,000 & \$405,000 & \$412,000 \\
Sale Price P99 & \$1,650,000 & \$1,680,000 & \$1,720,000 \\
Mean Assessed Value & \$30,059 & \$29,718 & \$30,213 \\
Assessed Value P1 & \$4,549 & \$4,499 & \$4,499 \\
Assessed Value P25 & \$15,243 & \$15,359 & \$15,669 \\
Assessed Value P50 & \$22,774 & \$22,799 & \$23,000 \\
Assessed Value P75 & \$34,079 & \$33,917 & \$34,091 \\
Assessed Value P99 & \$141,517 & \$137,475 & \$139,278 \\
Mean Ratio & 0.10 & 0.10 & 0.09 \\
Ratio P1 & 0.03 & 0.03 & 0.03 \\
Ratio P25 & 0.07 & 0.07 & 0.07 \\
Ratio P50 & 0.09 & 0.08 & 0.08 \\
Ratio P75 & 0.10 & 0.10 & 0.10 \\
Ratio P99 & 0.32 & 0.34 & 0.25 \\
\bottomrule
\end{tabular}

\caption{\textbf{Comparison of Sale Prices, Assessed Values, and Assessment Ratios Between Ground-Truth and Broker-Provided Data for Cook County, 2018--2021.} The table displays summary statistics (counts, means, and percentile values) for each of the Cook County (ground truth), ATTOM, and Cotality datasets for arms-length, single-family home transactions from 2018 to 2021. ``Ratio" indicates the ratio between the recorded assessed value and property sale price. ``P\#" indicate percentiles.}

\label{tab:summ_stats}
\end{table*}

\section{Coverage and Imputation Error in Brokered Data}\label{findings}
We identify two types of consequential errors in brokered data. First, we find evidence of \textit{imputation errors}. Imputation errors are issues of data accuracy, where records representing the same property transaction have different recorded sale prices in brokered data as compared to ground truth data, likely representing errors in how data brokers impute sale prices that are not directly available to them. Second, we find evidence of \textit{coverage errors}. Coverage errors are issues of data representativeness, attributable to missing data or differences in how deed and property characteristics are reported, such that transactions recorded as arms-length sales of single-family homes in ground truth data are not recorded as such in brokered data.\looseness=-1

\subsection{ATTOM} \label{attom}

We match transactions between ATTOM and ground-truth data to assess the coverage and accuracy of the ATTOM data. Our match process is described in Appendix \ref{app:match_procedure}.\looseness=-1

As shown in Table \ref{tab:attom_discrepancies_by_year}, we are able to match 134,427 ATTOM transactions to our ground-truth data, comprising 87.8\% of ground-truth sales. Among matched transactions, we find a total of 1,976 transactions that have discrepancies in sale price which exceed 5\% of the ground-truth price. As shown in Table \ref{tab:attom_ratio_counts}, the overwhelming majority of these discrepant transactions can be represented as simple ratios of broker-reported to ground-truth sale price. The most common ratios of ATTOM to ground-truth sale prices are $\frac{2}{3}$, $\frac{1}{3}$, and 2. While we also found instances of dropped, transposed, or otherwise mistranscribed digits,\footnote{Examples from the data: "\$7,000,000" instead of "\$700,000" (dropped digit), "\$200,000" instead of "\$220,000" (mistranscribed digit), and "\$271,000" instead of "\$217,000" (transposed digit).}  for the majority of cases, the discrepancies do not appear to be the result of data entry error.\looseness=-1

Imputation error can potentially explain the pattern of discrepancies we observe. In discussions with one of the two brokers examined here, we confirmed that when sale price is not directly available from county-level records sourced through FOIA requests, it may be imputed using local transfer tax rates. These transfer taxes are typically structured as a fixed percentage tax on a property's purchase price with de minimus exemptions. In Cook County, there are 75 distinct municipalities that independently set transfer tax rates.\footnote{A full list of transfer taxes across municipalities in Cook County as of January 2025 is available at \ttaxurl.} In addition to setting rates, municipalities may also divide responsibility for remitting the tax between the buyer and seller. In the event that a broker imputes a property's sale price using the incorrect municipality's transfer tax rate, or confuses the buyer's share of the tax for the seller's share, then the resulting imputed price will be an exact multiple of the actual price.\looseness=-1

\begin{table*}[ht]
\centering
\small
\adjustbox{max width=\textwidth}{%
\begin{tabular}{lcccc}
\toprule
 & \textbf{2018} & \textbf{2019} & \textbf{2020} & \textbf{2021} \\
\midrule
N matched & 32,656 & 31,935 & 34,315 & 35,521 \\
Matched as \% of ground-truth sales & 86.08\% & 86.91\% & 87.36\% & 90.90\% \\
\addlinespace
Share with sale price error $>$5\% & 1.27\% & 0.74\% & 3.70\% & 0.16\% \\
\quad Conditional mean error (\%) & 260.80\% & 79.33\% & 63.99\% & 78.97\% \\
  & (137.63\% - 383.97\%) & (65.29\% - 93.37\%) & (59.06\% - 68.92\%) & (44.28\% - 113.66\%) \\
\quad Conditional mean error (\$) & \$398,934 & \$166,999 & \$200,411 & \$162,269 \\
  & (\$245,412 - \$552,457) & (\$138,619 - \$195,380) & (\$187,999 - \$212,823) & (\$125,376 - \$199,162) \\
\addlinespace
Share with assessed value error $>$5\% & 0.02\% & 0.02\% & 0.06\% & 0.95\% \\
\quad Conditional mean error (\%) & 58.13\% & 35.54\% & 53.65\% & 13.57\% \\
  & (-26.69\% - 142.94\%) & (9.77\% - 61.30\%) & (-14.88\% - 122.19\%) & (12.73\% - 14.42\%) \\
\quad Conditional mean error (\$) & \$18,383 & \$13,300 & \$33,008 & \$4,292 \\
  & (\$4,273 - \$32,494) & (\$-3,588 - \$30,187) & (\$-19,923 - \$85,940) & (\$4,010 - \$4,573) \\
\bottomrule
\end{tabular}
}%
\caption{\textbf{Discrepancies between ATTOM and Ground Truth Sale Prices and Assessments in Cook County, 2018-2021.} The table displays summary statistics of the discrepancies between ATTOM and ground-truth transaction data for sales of single-family homes in Cook County from 2018–2021. “Error” is calculated as the absolute difference between the data broker and ground truth transaction price or assessed value as a fraction of the ground truth price or value. Conditional means are computed using only matched transactions with error $>$5\%. Records were matched using property address, latitude, longitude, and sale date. 95\% confidence intervals are displayed in parentheses.}
\label{tab:attom_discrepancies_by_year}
\begin{comment}
\vspace{4pt} % small vertical space
\begin{minipage}{\textwidth}
\scriptsize
\textit{Notes}: The table displays summary statistics of the discrepancies between ATTOM and ground-truth transaction data for sales of single-family homes in Cook County from 2018–2021. “Error” is calculated as the absolute difference between the data broker and ground truth transaction price or assessed value as a fraction of the ground truth price or value. Conditional means are computed using only matched transactions with error $>$5\%. Records were matched using property address, latitude, longitude, and sale date. 95\% confidence intervals are displayed in parentheses.
\end{minipage}
\end{comment}
\end{table*}

\setlength\tabcolsep{0.04\linewidth}
\begin{table}[h]
\small
\centering
\begin{tabular}{lrr}
\toprule
Ratio & Count & Share of Discrepancies (\%) \\
\midrule
0.667 ($2/3$) & 998 & 50.51 \\
2.000 & 152 & 7.69 \\
0.333 ($1/3$)& 116 & 5.87 \\
0.666 ($2/3$) & 95 & 4.81 \\
0.833 ($5/6$) & 28 & 1.42 \\
0.665 ($\approx 2/3$) & 18 & 0.91 \\
0.250 ($1/4$) & 15 & 0.76 \\
10.000 & 13 & 0.66 \\
1.999 ($\approx 2$) & 11 & 0.56 \\
1.133 ($17/15$) & 10 & 0.51 \\
Other (random) & 432 & 21.86 \\
Other (simple ratio) & 88 & 4.45 \\
Total & 1,976 & 100.00\% \\
\bottomrule
\end{tabular}

\caption{\textbf{Distribution of Ratios of ATTOM to Ground Truth Sale Prices for Cook County Discrepant Sales, 2018-2021.} The table shows the distribution of ratios of ATTOM-recorded sale prices to ground truth prices for matched discrepant sales in Cook County covering 2018-2021. A ``discrepant sale" is any sale where the absolute difference between the broker and ground-truth sale price exceeds 5\% of the ground-truth price.}
\label{tab:attom_ratio_counts}
\end{table}

An additional 371 matched ATTOM transactions have discrepancies in assessed value that exceed 5\% of ground-truth. These discrepancies do not follow a clear pattern, such as substitution between pre- and post-appeal assessed values. Just 6 instances are attributable to rounding error. Assessed value errors in ATTOM's data are far more limited in scope and magnitude than sale price errors.

We next turn to the portion of ground-truth transactions that we are unable to match to ATTOM's data, leading to coverage error between the two datasets.  Ground-truth transactions generally fail to match to brokered data for one of three reasons. First, brokers may misreport sale and property characteristics, labeling an arms-length transaction as a foreclosure or a single-family home as a condominium. Since we filter each dataset to include only arms-length transactions of single-family homes, any transaction with a misreported filter in brokered data will not match to its corresponding ground-truth sale. Next, a transaction may be completely missing or duplicated in the brokered data, in which case it will be dropped or missing prior to the match. Finally, transactions may not match due to the address and sale date similarity thresholds in our match procedure, which could exclude valid matches if too strict.\looseness=-1

Table \ref{tab:attom-mismatch-accounting} provides a breakdown of the causes of match failure for ATTOM. About half of match failures occur due to broker misreporting of sale and property characteristics. Another third are accounted for by missing records, while an additional 9\% are attributable to duplicate sales. The remaining 5\% of failed matches are largely due to the thresholds for address and sale month similarity in our matching process. We examine the robustness of our results to different choices of matching thresholds in Appendix \ref{app:match_procedure}.\looseness=-1

\begin{table}[ht]
\centering
\small
\adjustbox{max width=\textwidth}{%
\begin{tabular}{lr}
\textbf{N ground truth sales} & 153,044 \\ 
\textbf{N sales not matched to ATTOM} & 18,617 \\ 
\textbf{Reason for match failure:} &  \\ 
Record missing & 36.8\% \\ 
Matches duplicate sale & 8.6\% \\ 
Broker filter misreported & 49.8\% \\ 
\quad Arms-length sale & 38.0\% \\ 
\quad Not multiparcel & 2.6\% \\ 
\quad Sale price $>$ \$10k & 13.5\% \\ 
\quad Single-family home & 13.5\% \\ 
Does not meet fuzzy match criteria & 4.0\% \\ 
Other & 0.9\% \\ 
\hline 
\textbf{Total} & 100.0\% 
\end{tabular}
}%
\caption{\textbf{Sources of coverage error: ATTOM.} The table displays the sources of coverage error between ground truth and ATTOM data for arms-length sales of single family homes in Cook County from 2018-2021.}
\label{tab:attom-mismatch-accounting}
\begin{comment}
\vspace{4pt} % small vertical space
\begin{minipage}{\textwidth}
\scriptsize
\textit{Notes}: The table displays the sources of coverage error between ground truth and ATTOM data for arms-length sales of single family homes in Cook County from 2018-2021.
\end{minipage}
\end{comment}
\end{table}

\subsection{Cotality} \label{cotality}

Using the same match procedure applied above and described in Appendix \ref{app:match_procedure}, we are able to match 129,606 transactions between Cotality and Cook County, comprising 84.7\% of all ground-truth transactions. Among these transactions, we find 1,919 transactions where the sale price listed in Cotality differs from the ground-truth price by more than 5\%. Table \ref{tab:cotality_discrepancies_by_year} provides a summary of these discrepancies by year of sale.\looseness=-1

\begin{table*}[ht]
\label{tab:cotaltiy_discrepancies_by_year}
\centering
\small
\adjustbox{max width=\textwidth}{%
\begin{tabular}{lcccc}
\toprule
 & \textbf{2018} & \textbf{2019} & \textbf{2020} & \textbf{2021} \\
\midrule
N matched & 30,763 & 31,116 & 34,212 & 33,515 \\
Matched as \% of ground-truth sales & 81.09\% & 84.68\% & 87.10\% & 85.76\% \\
\addlinespace
Share with sale price error $>$5\% & 1.04\% & 0.60\% & 3.71\% & 0.43\% \\
\quad Conditional mean error (\%) & 94.40\% & 89.90\% & 67.78\% & 133.24\% \\
  & (78.32\% - 110.47\%) & (72.26\% - 107.54\%) & (62.11\% - 73.45\%) & (93.07\% - 173.41\%) \\
\quad Conditional mean error (\$) & \$215,846 & \$156,821 & \$203,642 & \$431,771 \\
  & (\$168,451 - \$263,242) & (\$127,478 - \$186,164) & (\$189,252 - \$218,032) & (\$286,907 - \$576,634) \\
\addlinespace
Share with assessed value error $>$5\% & 0.02\% & 0.00\% & 0.06\% & 0.97\% \\
\quad Conditional mean error (\%) & 67.20\% & 0.00\% & 54.48\% & 15.61\% \\
  & (-23.46\% - 157.87\%) & (0.00\% - 0.00\%) & (-13.88\% - 122.85\%) & (13.32\% - 17.90\%) \\
\quad Conditional mean error (\$) & \$18,979 & \$0 & \$30,792 & \$4,869 \\
  & (\$-7,826 - \$45,784) & (\$0 - \$0) & (\$-22,166 - \$83,749) & (\$4,268 - \$5,471) \\
\bottomrule
\end{tabular}
}%
\caption{\textbf{Discrepancies between Cotality and Ground Truth Sale Prices and Assessments in Cook County, 2018-2021.} The table displays summary statistics of the discrepancies between Cotality and ground-truth transaction data for sales of single-family homes in Cook County from 2018–2021. “Error” is calculated as the absolute difference between the data broker and ground truth transaction price or assessed value as a fraction of the ground truth price or value. Conditional means are computed using only matched transactions with error $>$5\%. Records were matched using property address, latitude, longitude, and sale date. 95\% confidence intervals are displayed in parentheses. Conditional means are not reported for cells with zero observations.}
\label{tab:cotality_discrepancies_by_year}
\begin{comment}
\vspace{4pt} % small vertical space
\begin{minipage}{\textwidth}
\scriptsize
\textit{Notes}: The table displays summary statistics of the discrepancies between Cotality and ground-truth transaction data for sales of single-family homes in Cook County from 2018–2021. “Error” is calculated as the absolute difference between the data broker and ground truth transaction price or assessed value as a fraction of the ground truth price or value. Conditional means are computed using only matched transactions with error $>$5\%. Records were matched using property address, latitude, longitude, and sale date. 95\% confidence intervals are displayed in parentheses.
\end{minipage}
\end{comment}
\end{table*}

Similar to ATTOM, we find that for the overwhelming majority of transactions with sale price discrepancies, the sale prices recorded by Cotality are exact multiples of ground-truth sale prices. And as with ATTOM, the most common ratios of Cotality to ground-truth sale prices are $\frac{2}{3}$, $\frac{1}{3}$, and $2$ as shown in Table \ref{tab:cotality_ratio_counts}, indicating that Cotality's data also plausibly suffers from imputation error.

\setlength\tabcolsep{0.04\linewidth}
\begin{table}[h]
\small
\centering
\begin{tabular}{lrr}
\toprule
Ratio & Count & Share of Discrepancies (\%) \\
\midrule
0.667 ($2/3$) & 1,093 & 56.96 \\
0.333 ($1/3$) & 138 & 7.19 \\
0.666 ($2/3$) & 100 & 5.21 \\
2.000 & 44 & 2.29 \\
0.833 ($5/6$) & 28 & 1.46 \\
0.250 ($1/4$)& 17 & 0.89 \\
0.095 & 14 & 0.73 \\
0.948 & 11 & 0.57 \\
0.665 ($\approx 2/3$) & 10 & 0.52 \\
1.133 ($17/15$) & 9 & 0.47 \\
Other (random) & 359 & 18.71 \\
Other (simple ratio) & 96 & 5.00 \\
Total & 1,919 & 100.00\% \\
\bottomrule
\end{tabular}

\caption{\textbf{Distribution of Ratios of Cotality to Ground Truth Sale Prices for Cook County Discrepant Sales, 2018-2021.} The table shows the distribution of ratios of Cotality-recorded sale prices to ground truth prices for matched discrepant sales in Cook County covering 2018-2021. A ``discrepant sale" is any sale where the absolute difference between the broker and ground-truth sale price exceeds 5\% of the ground-truth price.}
\label{tab:cotality_ratio_counts}
\begin{comment}
\vspace{4pt} % small vertical space
\begin{minipage}{\textwidth}
\scriptsize
      \scriptsize
      \textit{Notes}: The table shows the distribution of ratios of Cotality-recorded sale prices to ground truth prices for matched discrepant sales in Cook County covering 2018-2021. A ``discrepant sale" is any sale where the absolute difference between the broker and ground-truth sale price exceeds 5\% of the ground-truth price. 
    \end{minipage}
\end{comment}
\end{table}

An additional 352 matched Cotality transactions have discrepancies in assessed value that exceed 5\% of ground-truth. Similar to ATTOM's assessed value discrepancies, these discrepancies do not follow a clear pattern, such as substituting between pre- and post-appeal assessed values. Just 6 instances are attributable to rounding error. As with ATTOM, assessed value errors in Cotality's data are far more limited in scope and magnitude than sale price errors.

Table \ref{tab:cotality-mismatch-accounting} provides a breakdown of the causes of match failure for Cotality. In contrast to ATTOM, misreported sale and property characteristics explain a smaller share of match failures (32\% versus 50\%), while missing sale records explain comparatively larger share (56\% versus 37\%). Duplicate sales in brokered data explain about 9\% of match failures, and our choice of matching thresholds account for most of the remaining 3\% of unmatched transactions.

\begin{table}[ht]
\centering
\small
\adjustbox{max width=\textwidth}{%

\begin{tabular}{lr}
\textbf{N ground truth sales} & 153,044 \\ 
\textbf{N unmatched sales} & 23,438 \\ 
\textbf{Share of unmatched:} &  \\ 
Record missing & 56.0\% \\ 
Matches duplicate sale & 8.9\% \\ 
Broker filter misreported & 32.0\% \\ 
\quad Arms-length sale & 28.4\% \\ 
\quad Not multiparcel & 3.1\% \\ 
\quad Sale price $>$ \$10k & 9.0\% \\ 
\quad Single-family home & 9.0\% \\ 
Does not meet fuzzy match criteria & 2.3\% \\ 
Other & 0.7\% \\ 
\hline 
\textbf{Total} & 100.0\% 
\end{tabular}
}%
\caption{\textbf{Sources of coverage error: Cotality.} The table displays the sources of coverage error between ground truth and Cotality data for arms-length sales of single family homes in Cook County from 2018-2021.}
\label{tab:cotality-mismatch-accounting}
\begin{comment}
\vspace{4pt} % small vertical space
\begin{minipage}{\textwidth}
\scriptsize
\textit{Notes}: The table displays the sources of coverage error between ground truth and Cotality data for arms-length sales of single family homes in Cook County from 2018-2021. 
\end{minipage}
\end{comment}
\end{table}
\section{Shared Errors Across Brokers} \label{error-concordance}

One of the most common approaches to assessing robustness to data errors is a sensitivity analysis to different data sources \citep{hsu2024human, desai2019comparative, j2019geographic}. We therefore investigate whether reporting errors are correlated across sources at the transaction level, which would limit the diagnostic value of cross-source comparison, and find substantial dependence across data sources in errors for specific transactions. One reason for this dependence is that data brokers frequently sell and share data with one another, meaning errors can quickly propagate across networks of brokers \citep{FTCdatabrokers2014}. In the context of property data, antitrust concerns specifically resulted in an FTC consent degree that required Cotality (then-CoreLogic) to share data with ATTOM (then-RealtyTrac) \citep{lane2018}, and as such an analysis of shared errors is particularly appropriate.\looseness=-1

First, we consider transactions where broker and county-reported sale prices differ. Out of 1,976 discrepant ATTOM transactions and 1,919 discrepant Cotality transactions, we find that 1,512 transactions overlap\textemdash meaning brokers share roughly 3 out of every 4 discrepant transactions in common. Moreover, among these 1,512 shared discrepant transactions, the broker-reported sale prices are identical in virtually all cases (99.8\%).\looseness=-1

We next turn to coverage error, considering overlap between brokers in the ground truth transactions we are unable to match to their data, as well as the overlap in underlying reasons why such transactions fail to match. The results of this analysis are shown in Table \ref{tab:error_concordance}. ATTOM and Cotality share 11,094 unmatched transactions in common, representing roughly 50 to 60\% of each broker's total unmatched transactions. Among these transactions, the reasons for match failure are largely consistent across brokers. For example, one reason a transaction may fail to match is because it is missing in the broker's data. We find that, among the subset of unmatched transactions these brokers have in common, if a transaction is missing in Cotality's data, there is a 97\% chance it is also missing in ATTOM's data (5,432/5,615). We find similar agreement along the other dimensions we consider.\looseness=-1

\begin{table*}
\centering
\begin{adjustbox}{max width=\textwidth}
\begin{tabular}{ll|cccccc}
 & & \multicolumn{6}{c}{\textbf{ATTOM error type}} \\
 & & \textbf{Missing} & \textbf{Duplicate sale} & \textbf{Filter misreported} & \textbf{No fuzzy match} & \textbf{Other} & \textbf{Total} \\
\hline
\multirow{6}{*}{\shortstack{\textbf{Cotality} \\ \textbf{error type}}} & \textbf{Missing} & \textbf{5,432} & 150 & 625 & 40 & 4 & 6,251 \\
 & \textbf{Duplicate sale} & 24 & \textbf{810} & 290 & 137 & 1 & 1,262 \\
 & \textbf{Filter misreported} & 149 & 93 & \textbf{2,762} & 29 & 7 & 3,040 \\
 & \textbf{No fuzzy match} & 8 & 1 & 38 & \textbf{377} & 3 & 427 \\
 & \textbf{Other} & 2 & 1 & 8 & 2 & \textbf{101} & 114 \\
\hline
 & \textbf{Total} & 5,615 & 1,055 & 3,723 & 585 & 116 & 11,094 \\
\end{tabular}

\end{adjustbox}
\caption{\textbf{Match failure concordance matrix: Cotality vs. ATTOM.} The table shows the concordance in reasons for match failure between ATTOM and Cotality among unmatched ground-truth transactions the brokers have in common for Cook County from 2018-2021. ``Missing" indicates that the unmatched record is completley missing in the brokered dataset. ``Duplicate sale" indicates that record is duplicated in the brokered data and therefore dropped prior to matching. ``Filter misreported" indicates that one or more data filters (single-family home, arms-length transaction, sale price $>$ \$10K, or single-parcel sale) is misreported in the brokered data. "No fuzzy match" indicates that there is a potential match for the sale in the brokered data that does not meet our fuzzy match criteria for address and sale month similarity, as described in Appendix \ref{app:match_procedure}. Bolded values along the diagonal indicate cases where the reasons for match failure are the same across brokers.}
\label{tab:error_concordance}
\begin{comment}
\vspace{4pt} % small vertical space
\begin{minipage}{\textwidth}
\scriptsize
\textit{Notes}: The table shows the concordance in reasons for match failure between ATTOM and Cotality among unmatched ground-truth transactions the brokers have in common for Cook County from 2018-2021. ``Missing" indicates that the unmatched record is completley missing in the brokered dataset. ``Duplicate sale" indicates that record is duplicated in the brokered data and therefore dropped prior to matching. ``Filter misreported" indicates that one or more data filters (single-family home, arms-length transaction, sale price $>$ \$10K, or single-parcel sale) is misreported in the brokered data. "No fuzzy match" indicates that there is a potential match for the sale in the brokered data that does not meet our fuzzy match criteria for address and sale month similarity, as described in Appendix \ref{app:match_procedure}. Bolded values along the diagonal indicate cases where the reasons for match failure are the same across brokers.
\end{minipage}
\end{comment}
\end{table*}

Overall, we find considerable overlap in the errors present in ATTOM and Cotality's data. The most straightforward explanation for the similarities we observe in this particular case is the data sharing agreement between Cotality and ATTOM as mandated by the FTC's 2018 consent decree. However, this is not the only mechanism through which errors can propagate, and we cannot rule out other potential explanations. For example, these brokers may rely on similar imputation and processing methods. As previously noted, it is also common for brokers to sell data to one another \cite{FTCdatabrokers2014}; such transactions are rarely disclosed to end users and are subject to limited regulatory oversight \cite{sherman2023databrokerage}. These practices pose several challenges for researchers. First, researchers cannot use one broker's data to validate findings or correct errors in another broker's data if both datasets suffer from the same shortcomings. Second, combining or pooling brokered datasets to improve coverage or precision may yield smaller gains than expected if coverage gaps are largely shared. Finally, seemingly independent replication across studies using different brokers may provide a false sense of robustness, since convergent findings could reflect common errors rather than unbiased agreement.\looseness=-1
\section{Consequences of Data Errors} \label{implications}

Finally, we examine whether the coverage and imputation errors we identify above significantly affect estimates derived from brokered data using measures of property tax regressivity, a key economic statistic. Property taxes comprise a substantial share of revenues for local governments throughout the U.S. These taxes are allocated to households via property tax assessments, which are typically conducted at the county level and involve mapping property characteristics to market values using valuation tables or predictive models \citep{dornfestStateProvincialProperty}. While property taxes are meant to be proportional to each property's underlying market value, several studies have demonstrated that these taxes and corresponding assessments are in fact regressive: higher-valued properties tend to be under-assessed and have lower effective tax rates than lower-valued properties \citep{berry2021reassessing, amornsiripanitch2020residential, mcmillenAssessmentRegressivityProperty2020, smith2026tradeoffsdomaindependentimproving}.\looseness=-1

Studies of assessment regressivity typically compare assessed values to sale prices among homes that sell in a given year, under the assumption that these sale prices reflect the true underlying value of the home. These studies generally rely on data either obtained from counties through FOIA requests or compiled by data brokers such as Cotality and ATTOM \citep{berry2021reassessing, amornsiripanitch2020residential, avenancio2022assessment}. We estimate and compare standard regressivity measures from this literature using both broker-reported and ground-truth data. We consider the price-related differential (PRD) \citep{iaaoStandard2017}, the log coefficient \citep{international1978improving}, and the Suits Index \citep{suits1977measurement}. Descriptions of each measure are provided in Appendix \ref{appendix:regressivity-metrics}.\looseness=-1

\begin{figure*}
\centering
\includegraphics[width=\linewidth]{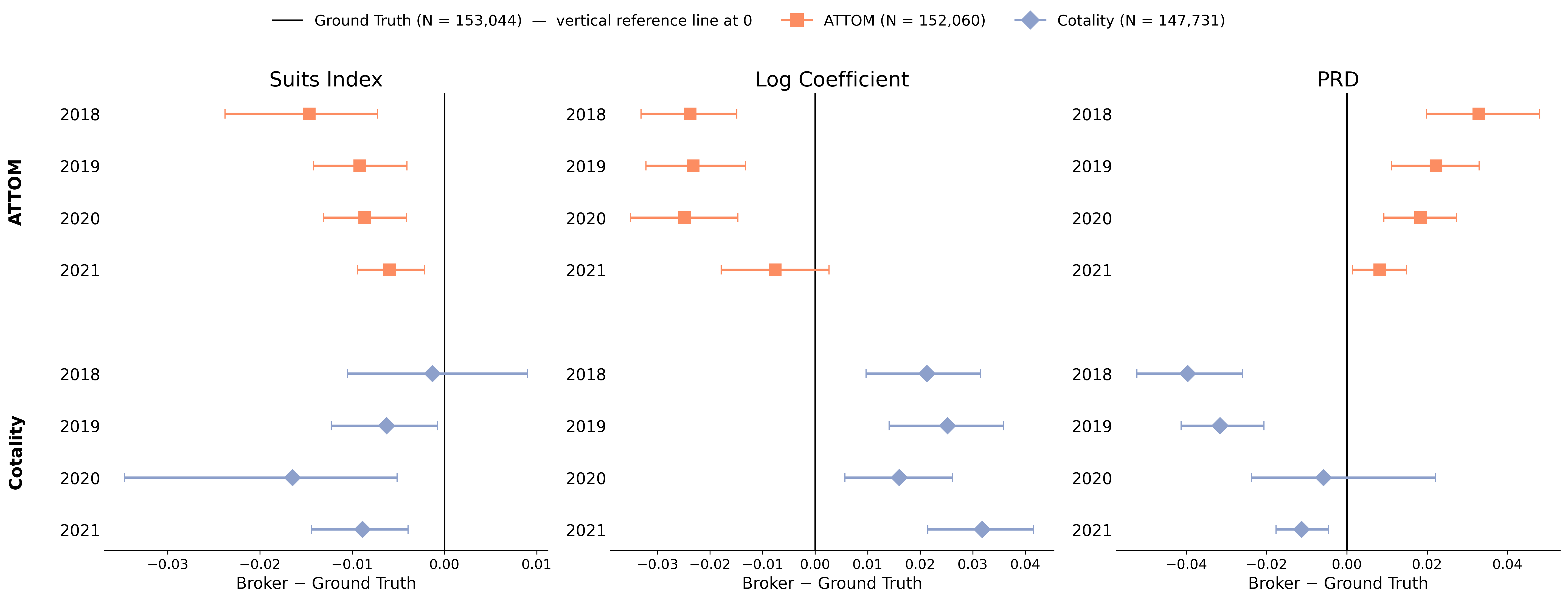}
\caption{\textbf{Difference between Ground Truth and Broker-Derived Regressivity Estimates by Year and Data Source, 2018-2021.} The figure shows the differences between estimates of regressivity derived from ground-truth administrative data and brokered data by year and data source for single-family home sales in Cook County covering 2018-2021. Ground truth values are derived from data published by the Cook County Assessor's office. 95\% confidence intervals are derived from a nonparametric percentile bootstrap (1,000 iterations), with each dataset resampled independently. For the Suits Index, lower values are more regressive. For the Log Coefficient, lower values are more regressive. For the PRD, higher values are more regressive.}
\label{fig:regressivity_metrics}
\end{figure*}

Our results, presented in Fig.~\ref{fig:regressivity_metrics}, show that regressivity estimates derived from broker-reported transaction data significantly differ from ground-truth values for at least one broker in all four sample years. In particular, Cotality data tend to underestimate the level of regressivity relative to ground truth data for the log coefficient and PRD measures, while ATTOM-derived estimates generally overstate regressivity for all three metrics. Differences between broker-reported and ground-truth estimates narrow by the end of our sample period.\looseness=-1

Both coverage and imputation error contribute to the gap between broker-reported and ground-truth measures of regressivity. To determine their relative contributions, we compare three estimates:\looseness=-1

\begin{itemize}
 \item a \textbf{ground-truth estimate}, computed from the full population of ground-truth data;
\item an \textbf{imputation-error estimate}, computed from the same ground-truth population but replacing ground-truth prices and assessments with broker-reported prices and assessments for matched transactions; and
\item a \textbf{broker estimate}, computed from the full population of broker-reported data.
\end{itemize}

The difference between the imputation-error and ground-truth estimates approximates the contribution of imputation error, while the difference between the broker and imputation-error estimates approximates the contribution of coverage error. The results of this analysis are provided in Tables \ref{tab:attom-error-decomp} and \ref{tab:cotality-error-decomp}.\looseness=-1

\begin{table*}[ht]
\centering
\small
\adjustbox{max width=0.53\textwidth}{%
\begin{tabular}{llccc}
& \multicolumn{4}{c}{\textbf{Panel A: Differences in Levels}}\\
\toprule
 & Metric & Total Difference & Imputation Error & Coverage Error \\
\midrule
2018 & Log Coefficient & 0.020$^{*}$ & 0.008 & 0.012 \\
 & PRD & -0.026$^{*}$ & -0.012 & -0.015 \\
 & Suits Index & 0.009$^{*}$ & 0.008 & 0.001 \\
2019 & Log Coefficient & 0.017$^{*}$ & 0.002 & 0.015 \\
 & PRD & -0.015$^{*}$ & -0.002 & -0.013 \\
 & Suits Index & 0.003$^{*}$ & 0.001 & 0.003 \\
2020 & Log Coefficient & 0.022$^{*}$ & 0.013 & 0.009 \\
 & PRD & -0.015$^{*}$ & -0.010 & -0.005 \\
 & Suits Index & 0.006$^{*}$ & 0.006 & -0.000 \\
2021 & Log Coefficient & 0.003 & 0.003 & -0.001 \\
 & PRD & -0.004$^{*}$ & -0.003 & -0.001 \\
 & Suits Index & 0.002$^{*}$ & 0.000 & 0.001 \\
\bottomrule
\end{tabular}
}
\adjustbox{max width=0.44\textwidth}{%
\begin{tabular}{llcc}
& \multicolumn{3}{c}{\textbf{Panel B: Share of Total Difference (Absolute Contributions)}}\\
\toprule
 & Metric & Imputation Error & Coverage Error \\
\midrule
2018 & Log Coefficient & 39.93\% & 60.07\% \\
 & PRD & 43.5\% & 56.5\% \\
 & Suits Index & 88.28\% & 11.72\% \\
2019 & Log Coefficient & 11.96\% & 88.04\% \\
 & PRD & 11.76\% & 88.24\% \\
 & Suits Index & 23.41\% & 76.59\% \\
2020 & Log Coefficient & 60.09\% & 39.91\% \\
 & PRD & 65.54\% & 34.46\% \\
 & Suits Index & 99.48\% & 0.52\% \\
2021 & Log Coefficient & 85.2\% & 14.8\% \\
 & PRD & 83.66\% & 16.34\% \\
 & Suits Index & 26.97\% & 73.03\% \\
\bottomrule
\end{tabular}
}%

\caption{\textbf{Decomposition of Differences Between Ground Truth and Broker Estimates: ATTOM.} Panels A and B decompose the relative contribution of imputation and coverage error to the observed differences between ATTOM and ground-truth regressivity estimates. We derive the values in each panel using the ground-truth, imputation error, and broker estimates of regressivity derived in Section \ref{implications}. Panel A shows the overall difference between ground truth and ATTOM estimates for each metric. Asterisks indicate statistical significance at the 5\% level. Panel B shows the absolute percentage contribution of Imputation and Coverage error to the total differences displayed in Panel A. ``Absolute percentage contribution" is measured as $|error\_type| / (|imputation\_error|+|coverage\_error|)$.}
\label{tab:attom-error-decomp}
\end{table*}

\begin{table*}
\centering
\small
\adjustbox{max width=0.53\textwidth}{%
\begin{tabular}{llccc}
& \multicolumn{4}{c}{\textbf{Panel A: Differences in Levels}}\\
\toprule
 & Metric & Total Difference & Imputation Error & Coverage Error \\
\midrule
2018 & Log Coefficient & -0.025$^{*}$ & 0.003 & -0.028 \\
 & PRD & 0.046$^{*}$ & -0.003 & 0.049 \\
 & Suits Index & -0.004 & 0.002 & -0.006 \\
2019 & Log Coefficient & -0.031$^{*}$ & 0.002 & -0.033 \\
 & PRD & 0.038$^{*}$ & -0.001 & 0.040 \\
 & Suits Index & 0.001$^{*}$ & 0.000 & 0.000 \\
2020 & Log Coefficient & -0.019$^{*}$ & 0.011 & -0.030 \\
 & PRD & 0.009 & -0.008 & 0.017 \\
 & Suits Index & 0.014$^{*}$ & 0.006 & 0.008 \\
2021 & Log Coefficient & -0.037$^{*}$ & 0.005 & -0.042 \\
 & PRD & 0.016$^{*}$ & -0.003 & 0.019 \\
 & Suits Index & 0.005$^{*}$ & 0.003 & 0.002 \\
\bottomrule
\end{tabular}
}
\adjustbox{max width=0.44\textwidth}{%
\begin{tabular}{llcc}
& \multicolumn{3}{c}{\textbf{Panel B: Share of Total Difference (Absolute Contributions)}}\\
\toprule
 & Metric & Imputation Error & Coverage Error \\
\midrule
2018 & Log Coefficient & 8.89\% & 91.11\% \\
 & PRD & 5.33\% & 94.67\% \\
 & Suits Index & 24.3\% & 75.7\% \\
2019 & Log Coefficient & 4.91\% & 95.09\% \\
 & PRD & 3.0\% & 97.0\% \\
 & Suits Index & 85.43\% & 14.57\% \\
2020 & Log Coefficient & 26.67\% & 73.33\% \\
 & PRD & 31.58\% & 68.42\% \\
 & Suits Index & 42.64\% & 57.36\% \\
2021 & Log Coefficient & 10.74\% & 89.26\% \\
 & PRD & 13.62\% & 86.38\% \\
 & Suits Index & 59.12\% & 40.88\% \\
\bottomrule
\end{tabular}
}%

\caption{\textbf{Decomposition of Differences Between Ground Truth and Broker Estimates: Cotality.} Panels A and B decompose the relative contribution of imputation and coverage error to the observed differences between Cotality and ground-truth regressivity estimates. We derive the values in each panel using the ground-truth, imputation error, and broker estimates of regressivity derived in Section \ref{implications}. Panel A shows the overall difference between ground truth and Cotality estimates for each metric. Asterisks indicate statistical significance at the 5\% level. Panel B shows the absolute percentage contribution of Imputation and Coverage error to the total differences displayed in Panel A. ``Absolute percentage contribution" is measured as $|error\_type| / (|imputation\_error|+|coverage\_error|)$.}
\label{tab:cotality-error-decomp}
\begin{comment}
\vspace{4pt} % small vertical space
\begin{minipage}{\textwidth}
\scriptsize
\textit{Notes}: Panels A and B decompose the relative contribution of imputation and coverage error to the observed differences between Cotality and ground-truth regressivity estimates. We derive the values in each panel using the ground-truth, imputation error, and broker estimates of regressivity derived in Section \ref{implications}. Panel A shows the overall difference between ground truth and Cotality estimates for each metric. Panel B shows the absolute percentage contribution of Imputation and Coverage error to the total differences displayed in Panel A. ``Absolute percentage contribution" is measured as $|error\_type| / (|imputation\_error|+|coverage\_error|)$
\end{minipage}
\end{comment}
\end{table*}

We observe key differences in the relative contributions of coverage and imputation error across brokers. For ATTOM, coverage error dominates in 2018 and 2019, when differences between its estimates and ground-truth estimates are statistically significant and the rate of imputation error in the data is relatively low ($\sim$1\%). In 2020, the share of transactions suffering from imputation error jumps to 3.7\%, and imputation error accounts for a larger portion of the overall difference in estimates. Finally, in 2021, the differences between ATTOM and ground-truth estimates more or less disappear, and the relative shares of imputation and coverage error reflect division of near-zero differences.\looseness=-1

Cotality's data, by contrast, is susceptible to a higher degree of coverage error, as discussed in Section \ref{cotality}. Correspondingly, coverage error predominates in virtually all cases where the differences between ground truth and Cotality estimates are statistically significant. We observe a slight increase in the share of error explained by imputation in 2020, where Cotality's data\textemdash similar to ATTOM's data\textemdash exhibits a temporary spike in the overall share of transactions suffering from imputation error (from ~1\% to 3.7\%). But overall, it is the difference in the population of arms-length transactions rather than difference in reported prices that drives most of the differences between Cotality and ground-truth estimates.\looseness=-1

The magnitude of estimated assessment regressivity differs significantly across brokered data sources, and the direction of bias is broker-specific. Decomposing the discrepancies into their underlying sources indicates that imputation error introduces a consistent, if modest, attenuation bias: both brokers' data initially understate the true degree of regressivity. Coverage error, by contrast, varies in sign and magnitude across brokers and years, and it dominates the total discrepancy for Cotality throughout the sample period and for ATTOM in earlier years. While ATTOM and Cotality share roughly half of their unmatched transactions, the remaining portion is large and idiosyncratic enough to drive substantial wedges between their regressivity estimates. The substantial overlap in errors documented in Section \ref{error-concordance} therefore does not translate into agreement at the level of regressivity estimates.\looseness=-1

\section{Discussion}\label{conclusion}

Brokered data now comprise a substantial share of the evidence base for academic and government research and the training data for machine learning models, including models deployed in the public sector. The credibility of these endeavors crucially depends on the brokers who source, process, and publish this data, and who generally treat their data lineage as proprietary and confidential. This places users of brokered data in a difficult position. On the one hand, rising expectations\textemdash for the scope and coverage of contemporary empirical research, or the technical capacity of government\textemdash encourage reliance on brokered data. On the other hand, users cannot account for how these data were sourced and processed, and they often have no practical way to validate that the data are as accurate or comprehensive as brokers claim.\looseness=-1

In this paper, we audit two central brokered property datasets, documenting significant shortcomings in their representativeness and accuracy. We find similar patterns of errors in each dataset and a high rates of concordance between brokers in the data reporting errors they make for specific transactions. While errors in sale price and assessed value tend to attenuate regressivity estimates, errors in data coverage bias estimates in unpredictable ways, and are responsible for a much larger portion of the bias we observe.\looseness=-1

The types of errors our audit identifies\textemdash nonrandom label noise and nonrandom sample selection\textemdash pose particular challenges for machine learning practitioners. First, noisy labels can degrade model accuracy, increase model complexity, and increase the number of training samples required to achieve reliable performance \cite{frenay2013classification}. Straightforward approaches to filtering errors, such as censoring or winsorizing extreme values, are unlikely to address the data issues we identify, as the vast majority of discrepant sales in both Cotality and ATTOM's datasets are not outliers with respect to sale price, assessed value, or assessment ratios. Furthermore, unrepresentative samples can bias model predictions \cite{zadrozny2004learning}. Practitioners can of course re-weight brokered data to better reflect the population of interest; however, doing so requires access to population-level characteristics, and researchers may not think to take this step if they believe their data are already comprehensive.\looseness=-1

Our findings underscore the importance of accessible open data from federal, state, and local governments. Federal governments, in particular, are the primary alternative to data brokers in supplying large-scale data on topics including population dynamics, economic activity, healthcare, and the environment. In contrast to data brokers, governments do not face competitive pressure to keep their data lineage closely held, and are in fact generally required to publicly disclose how data are sampled and processed \citep{eurostat_code_of_practice_2018, omb_spd1_2014, omb_statistical_policy_directive_4_2008}. Researchers equipped with these disclosures can more easily identify and account for imperfections in government data as compared to brokered data, whose provenance is opaque to them. Even smaller-scale open data, like the county-level data we consider here, provide researchers the means to validate the robustness and accuracy of their findings, and in certain settings can enable researchers to adjust brokered data for measurement and sampling error using sampling weights, instrumental variables, or other statistical tools. Brokers, of course, could also disclose their data lineage so that researchers have a better understanding of how their evidence is sourced and processed prior to analysis. 

What means do individuals and institutions have to correct errors in brokered data and redress harms caused by the application of such data in private and public sector systems? In California, individuals can compel brokers to fix errors and delete data upon request \citep{CCPA}. However, for an individual to exercise these rights, they must know that brokers have collected data about them or their property, know that this data is incorrect, and know that they have the right to correct errors or request deletion. While individuals may request that brokers share data they have collected about them and their properties under California and other state laws, these individuals cannot fully understand the impact of flaws in these brokers' data on their property's appraisal or tax bill without far more substantial visibility over brokered datasets and how appraisers and assessors use such data. Furthermore, for such laws to be effective, brokers must act on requests for data to be modified or deleted. In a study of data brokers registered in the state of California, \citet{vankempen2025consumerbewareexploringdata} find that roughly half of brokers do not respond at all to such requests \citep{gueorguieva2026}.\looseness=-1

Courts and administrative agencies can provide relief for egregious harms. For example, in 2025, the Federal Trade Commission issued a consent agreement addressing General Motors' collection and sale of drivers' precise geolocation data to consumer reporting agencies without drivers' express consent. The agreement ordered GM and related parties to delete the data and to stop collecting similar data in the future without clear, affirmative consent from drivers. This consent agreement had positive spillovers beyond GM in protecting consumers' right to privacy. However, such agreements are exceptional. They occur after harm has been done, and they cannot substitute for effective disclosure requirements regarding brokered data provenance and other ex ante consumer protections.\looseness=-1

Overall, our findings underscore the importance of broker transparency regarding the conceptual and statistical methods they use for data gathering and imputation. In the absence of such transparency, end users are left in the dark regarding potential shortcomings of their data, and brokers miss out on opportunities to improve data quality. Based on our findings, we recommend that researchers and practitioners evaluate the robustness of their estimates and model performance using open, ground-truth administrative data where available. We also encourage clear standards of transparency for data brokers regarding data provenance.\looseness=-1

\paragraph{Limitations.} We note several limitations of our work. First, our analysis focuses primarily on Cook County, IL. This is because Cook County is an outlier in terms of the amount and quality of property data that it makes available for researchers to download and use. To address this, we provide robustness checks for two other major U.S. cities with similar, albeit slightly less detailed, open data in Appendix~\ref{appendix:nyc}, showing that coverage and imputation errors also arise in those cities. Second, although we treat county-provided data as ``ground truth'', we note that, in a very small minority of cases, county-provided data contain errors that brokered data do not (\S\ref{data}). Although we believe that these instances are rare enough that they do not materially affect our analysis, we note that future audits of brokered data should continue to validate county-provided data against administrative records directly. Finally, without access to data brokers' documentation of their data provenance and processing steps, we cannot definitively verify our hypotheses about the potential causes of the coverage and imputation error that we identify.\looseness=-1

\section{Conclusion}
Through an audit of two prominent brokered property datasets against county-provided ``ground truth'' data on sales of single-family homes from 2018 to 2021 in Cook County, IL, we find evidence of coverage and imputation errors which bias key measures of economic inequality. Coverage error, which affects 12 to 15\% of transactions, occurs when missing data and conceptual differences in the reporting of deed and property characteristics cause transactions recorded as arms-length sales of single-family homes in ground truth data are not recorded as such in brokered data. Imputation error, which affects 1-2\% of matched transactions, occurs when records representing the same property transaction have different recorded sale prices in brokered data as compared to ground truth data, likely representing errors in how data brokers impute sale prices that are not directly available to them. We show that misreporting is highly consistent between brokers, suggesting that researchers cannot use one broker's data to validate findings or correct errors in another broker's data. Our findings highlight the crucial importance of open administrative data and transparency from brokers regarding data provenance and lineage.

\newpage

\section*{Ethical Considerations}
Although all of our data are based on public records, brokered data do include personally identifiable information, including the names and contact information of individuals involved in property transactions. In order to respect the privacy of those individuals, we did not use those fields in our analysis; further, we accessed brokered data only via a secure server.

\section*{Adverse Impacts}
The primary potential adverse impact of this work is that it runs the risk of providing ``free labor'' for data brokers by identifying issues in their data on their behalf so that they can be corrected. While we do acknowledge this as a risk, our hope is that this work instead draws attention to errors in brokered data, encouraging further audits, more transparency from brokers, and the wider release of administrative data.

\bibliography{references}

\newpage
\appendix
\setcounter{table}{0}
\renewcommand{\thetable}{A\arabic{table}}
\clearpage

\begingroup % localize the following settings                                                                    
\setlength\tabcolsep{2pt}
\footnotesize

\setlength\LTcapwidth{\textwidth} % default: 4in (rather less than \textwidth...)      

\setlength\LTleft{0pt}            % default: \fill
\setlength\LTright{0pt}           % default: \fill       

\onecolumn
\section{Prior Work Using Brokered Housing Data}
\captionsetup{font=normalsize}
\begin{longtable}{m{0.2\linewidth} m{0.3\linewidth} m{0.45\linewidth}}

\caption{Prior Work Using Brokered Data}
\label{tab:cotality_papers}
\\
\toprule
\multicolumn{1}{c}{\textbf{Paper}} & 
\multicolumn{1}{c}{\textbf{Brokered Data Source}} & 
\multicolumn{1}{c}{\textbf{Summary}} \\
\midrule
\endfirsthead

\toprule
\multicolumn{1}{c}{\textbf{Paper}} & 
\multicolumn{1}{c}{\textbf{Brokered Data Source}} & 
\multicolumn{1}{c}{\textbf{Summary}} \\
\endhead

\cite{amornsiripanitch2020residential} & Cotality Owner Transfer and Tax Data, 2000--2019 & Nationwide study of property tax assessment regressivity in the U.S. \\
\hline

\cite{phillyfedclimate} & Cotality Climate Risk Data, 2021 & Describes the geographic and demographic distribution of climate-related property risks in the U.S. \\
\hline

\cite{an2024capitalization} & Cotality Owner Transfer and Tax Data, 1980--2023 & Estimates capitalization of property taxes into home prices in Philadelphia. \\
\hline

\cite{avenancio2022assessment} & ATTOM Tax and Transaction Data, 2003-2016 & Nationwide study examining racial differences in residential property tax assessments. \\
\hline

%\cite{berry2024jure} & First American Tax and Sale Price Data, 2020--2022 & Compares statutory and de-facto property tax rates across 74 U.S. cities. \\
%\hline

\cite{berry2021reassessing} & Cotality Owner Transfer and Tax Data, 2006--2016 & Nationwide study of property tax assessment regressivity in the U.S. \\
\hline

\cite{diamond2024racial} & Cotality Owner Transfer and MLS Data, 2007--2017 & National study of racial differences in the rate of return on owner-occupied housing in the U.S. \\
\hline

\cite{ding2020effects} & Cotality Property and Owner Transfer Data, 2010--2018 & Measures impact of gentrification on homeowner mobility and tax delinquency in Philadelphia. \\
\hline

\cite{dallasfedclimate} & Cotality Climate Risk Data & Analyzes impact of insurance premiums on mortgage outcomes and buyer creditworthiness. \\
\hline

\cite{heilbron2024} & Cotality Climate Risk, Tax, and Mortgage Data, 2020--2022 & Describes the geographic and demographic distribution of climate-related property risks in the U.S. \\
\hline

\cite{horton2024property} & Cotality Owner Transfer, Tax, Involuntary Liens, and Building Permits Data, 2015--2019 & Analyses how property tax relief measures affect home prices and housing affordability in the U.S. \\
\hline

\cite{kim2023does} & Cotality Owner Transfer Data, 2001--2020 & Estimates effect of climate risk on migration and gentrification for hurricane-prone coastal cities in the U.S. \\
\hline

\cite{makridis2019tracking} & Cotality Property Data, 2006--2011 & Tracks migration of households following foreclosure. \\
\hline

\cite{mayer2022impact} & Cotality Owner Transfer Data, 2015--2020 & Estimates impact of “Opportunity Zone” development incentives on housing prices across the U.S. \\
\hline

\cite{mcmillen2020assessment} & Cotality Property and Owner Transfer Data, 2014--2016 & Study of assessment regressivity for 4 urban U.S. counties. \\
\hline

\cite{mcmillenMeasuresVerticalInequality2023} & Cotality Property and Owner Transfer Data (years unavailable) & Compares assessment regressivity measures across 48 large U.S. cities. \\
\hline

\cite{gao2012} & Cotality Mortgage Data, 2007--2011 & GAO report assessing and identifying opportunities to improve government response to the U.S. housing crisis. \\
\hline

\cite{gao2011} & Cotality Mortgage Data, 2001--2010 & GAO report assessing the effect of the Dodd-Frank act on the availability and affordability of mortgages. \\
\hline

\cite{skilling_deeper_2015} & Cotality Owner Transfer and Mortgage Data, 2013--2015 & Examines housing market conditions in New Zealand following introduction of loan-to-value lending constraints. \\
\hline

\cite{thomson2023identifying} & Cotality Property Data, 2019 & Estimates economic value of heirs' property across 11 southern and Appalachian U.S. states. \\

%https://www.huduser.gov/portal/periodicals/cityscpe/vol21num2/ch8.pdf#:~:text=households%20that%20experience%20foreclosure%20from,proceed%20to%20describe%20the%20nature
\bottomrule
\end{longtable}
\twocolumn

\endgroup
\section{Data Discrepancies Beyond Cook County}\label{appendix:nyc}

To assess whether the discrepancies we observe are limited to a specific geography, we compare broker-reported transactions to open data from New York City and Philadelphia. 

\paragraph{New York City.}
We matched records of single-family homes, condominiums, and duplexes from brokered data to ground-truth records from New York County (Manhattan), Bronx County, Kings County (Brooklyn), Queens County, and Richmond County (Staten Island), downloaded from NYC Open Data, for transactions conducted in 2018.\footnote{We were not able to find publicly available assessment data in NYC for the years 2019--2021.} We followed an analogous matching procedure to that described in Appendix~\ref{app:match_procedure}, with some modifications as we were not able to identify a flag for arms-length or multi-parcel transactions. A summary of NYC data is shown in Appendix Table~\ref{tab:summ_stats_nyc}.\looseness=-1

\paragraph{Philadelphia.}
We matched records of single-family homes from brokered data to ground-truth records from Philadelphia County, downloaded from Open Data Philly, for transactions conducted from 2018 to 2021.\footnote{\url{https://opendataphilly.org/}} We followed an analogous matching procedure to that described in Appendix~\ref{app:match_procedure}, with some modifications. As with the New York City open data,  we were not able to identify a flag for arms-length or multi-parcel transactions. A summary of NYC data is shown in Appendix Table~\ref{tab:summ_stats_philly}.\looseness=-1

\paragraph{Findings.} These data are not as complete as the data available from Cook County: we were not able to identify flags for arms-length or multi-parcel transactions in either dataset, and we were only able to find assessment data from 2018 for New York City. In both cases, we are able to match a lower percentage of transactions and observe a higher incidence of assessment error as compared to the Cook County analysis; this may be due to differences in address formatting in New York City and Philadelphia that makes transactions somewhat more challenging to match to brokered data as well as assessment data that is calculated differently across different cities and that may not reflect post-appeal values.\footnote{For these reasons, we do not calculate sources of coverage error for either city. We also note that we observe an unexpectedly high number of properties assessed as zero valued in ATTOM for Philadelphia, which skews estimates of assessment error between Open Data Philly and ATTOM; we are not able to determine why this occurs.} Nevertheless, we identify the presence of potential coverage error (i.e., unmatched properties) and imputation error (i.e., simple ratios of ground truth to brokered sale price) in both cities. Notably, for matched properties in both cities, we observe frequent, small discrepancies in assessed values and (as in Cook County) rare but very large discrepancies in sale prices (Appendix Tables~\ref{tab:attom_discrepancies_by_year_nyc}, ~\ref{tab:cotality_discrepancies_by_year_nyc},  ~\ref{tab:attom_discrepancies_by_year_philly}, ~\ref{tab:cotality_discrepancies_by_year_philly}). These discrepancies appear more likely to be transcription errors (i.e., they often are missing a zero or else have an extra zero appended), but we also identified the presence of other simple ratios that, like Cook County, appear likely to be imputation errors (Appendix Tables~\ref{tab:attom_ratio_counts_nyc}, ~\ref{tab:cotality_ratio_counts_nyc},  ~\ref{tab:attom_ratio_counts_philly}, ~\ref{tab:cotality_ratio_counts_philly}). \looseness=-1

\begin{table*}
\centering
\small
\begin{tabular}{llll}
\toprule
 & Ground Truth & ATTOM & Cotality \\
\midrule
N & 11,794 & 10,518 & 10,845 \\
Mean Sale Price & \$750,025 & \$740,398 & \$751,676 \\
Sale Price P1 & \$125,000 & \$120,219 & \$147,760 \\
Sale Price P25 & \$441,821 & \$449,063 & \$450,000 \\
Sale Price P50 & \$575,000 & \$577,794 & \$580,000 \\
Sale Price P75 & \$800,000 & \$800,000 & \$800,000 \\
Sale Price P99 & \$3,877,536 & \$3,700,000 & \$4,044,600 \\
Mean Assessed Value & \$663,983 & \$497,079 & \$594,448 \\
Assessed Value P1 & \$173,930 & \$138,398 & \$182,867 \\
Assessed Value P25 & \$396,000 & \$343,238 & \$379,000 \\
Assessed Value P50 & \$481,000 & \$416,200 & \$476,000 \\
Assessed Value P75 & \$692,000 & \$534,208 & \$621,000 \\
Assessed Value P99 & \$3,304,190 & \$1,708,456 & \$2,595,240 \\
Mean Ratio & 1.08 & 0.87 & 0.93 \\
Ratio P1 & 0.25 & 0.22 & 0.30 \\
Ratio P25 & 0.76 & 0.61 & 0.68 \\
Ratio P50 & 0.86 & 0.72 & 0.83 \\
Ratio P75 & 1.00 & 0.85 & 0.99 \\
Ratio P99 & 3.78 & 3.15 & 2.98 \\
\bottomrule
\end{tabular}
\caption{\textbf{Comparison of Sale Prices, Assessed Values, and Assessment Ratios Between Ground-Truth and Broker-Provided Data for New York City, 2018.} The table displays summary statistics (counts, means, and percentile values) for each of the NYC (ground truth), ATTOM, and Cotality datasets for arms-length, single-family home transactions in 2018. ``Ratio" indicates the ratio between the recorded assessed value and property sale price.}
\label{tab:summ_stats_nyc}
\end{table*}

\begin{table*}
\centering
\small
\begin{tabular}{lc}
\toprule
 & \textbf{2018} \\
\midrule
N matched & 8,184 \\
Matched as \% of ground-truth sales & 69.39\% \\
\addlinespace
Share with sale price error $>$5\% & 0.88\% \\
\quad Conditional mean error (\%) & 99.11\% \\
  & (63.57\% - 134.66\%) \\
\quad Conditional mean error (\$) & \$772,216 \\
  & (\$451,989 - \$1,092,444) \\
\addlinespace
Share with assessed value error $>$5\% & 83.15\% \\
\quad Conditional mean error (\%) & 23.46\% \\
  & (22.81\% - 24.12\%) \\
\quad Conditional mean error (\$) & \$164,684 \\
  & (\$155,547 - \$173,820) \\
\bottomrule
\end{tabular}
\caption{\textbf{Discrepancies between ATTOM and Ground Truth Sale Prices and Assessments in New York City, 2018.} The table displays summary statistics of the discrepancies between ATTOM and ground-truth transaction data for sales of single-family homes in New York City in 2018. “Error” is calculated as the absolute difference between the data broker and ground truth transaction price or assessed value as a fraction of the ground truth price or value. Conditional means are computed using only matched transactions with error $>$5\%. Records were matched using property address, latitude, longitude, and sale date. 95\% confidence intervals are displayed in parentheses.}
\label{tab:attom_discrepancies_by_year_nyc}
\end{table*}

\begin{table}
\centering
\small
\begin{tabular}{lrr}
\toprule
Ratio & Count & Share of Discrepancies (\%) \\
\midrule
10.000 & 12 & 16.67 \\
0.286 & 3 & 4.17 \\
0.281 & 3 & 4.17 \\
0.100 & 2 & 2.78 \\
0.400 & 2 & 2.78 \\
1.333 & 2 & 2.78 \\
2.802 & 1 & 1.39 \\
2.000 & 1 & 1.39 \\
2.009 & 1 & 1.39 \\
2.042 & 1 & 1.39 \\
Other (random) & 41 & 56.94 \\
Other (simple ratio) & 3 & 4.17 \\
Total & 72 & 100.00\% \\
\bottomrule
\end{tabular}
\caption{\textbf{Distribution of Ratios of ATTOM to Ground Truth Sale Prices for New York City Discrepant Sales, 2018.} The table shows the distribution of ratios of ATTOM-recorded sale prices to ground truth prices for matched discrepant sales in New York City in 2018. A ``discrepant sale" is any sale where the absolute difference between the broker and ground-truth sale price exceeds 5\% of the ground-truth price.}
\label{tab:attom_ratio_counts_nyc}
\end{table}

\begin{table*}
\small
\centering
\begin{tabular}{lc}
\toprule
 & \textbf{2018} \\
\midrule
N matched & 8,465 \\
Matched as \% of ground-truth sales & 71.77\% \\
\addlinespace
Share with sale price error $>$5\% & 0.72\% \\
\quad Conditional mean error (\%) & 122.20\% \\
  & (72.46\% - 171.94\%) \\
\quad Conditional mean error (\$) & \$996,995 \\
  & (\$527,431 - \$1,466,559) \\
\addlinespace
Share with assessed value error $>$5\% & 85.17\% \\
\quad Conditional mean error (\%) & 21.93\% \\
  & (21.18\% - 22.68\%) \\
\quad Conditional mean error (\$) & \$134,556 \\
  & (\$126,747 - \$142,365) \\
\bottomrule
\end{tabular}
\caption{\textbf{Discrepancies between Cotality and Ground Truth Sale Prices and Assessments in New York City, 2018.} The table displays summary statistics of the discrepancies between Cotality and ground-truth transaction data for sales of single-family homes in New York City in 2018. “Error” is calculated as the absolute difference between the data broker and ground truth transaction price or assessed value as a fraction of the ground truth price or value. Conditional means are computed using only matched transactions with error $>$5\%. Records were matched using property address, latitude, longitude, and sale date. 95\% confidence intervals are displayed in parentheses.}
\label{tab:cotality_discrepancies_by_year_nyc}
\end{table*}

\begin{table}
\centering
\small
\begin{tabular}{lrr}
\toprule
Ratio & Count & Share of Discrepancies (\%) \\
\midrule
10.000 & 6 & 9.84 \\
0.286 & 4 & 6.56 \\
0.100 & 3 & 4.92 \\
0.281 & 3 & 4.92 \\
0.400 & 3 & 4.92 \\
2.176 & 1 & 1.64 \\
1.442 & 1 & 1.64 \\
1.461 & 1 & 1.64 \\
1.488 & 1 & 1.64 \\
1.522 & 1 & 1.64 \\
Other (random) & 31 & 50.82 \\
Other (simple ratio) & 6 & 9.84 \\
Total & 61 & 100.00\% \\
\bottomrule
\end{tabular}
\caption{\textbf{Distribution of Ratios of Cotality to Ground Truth Sale Prices for New York City Discrepant Sales, 2018.} The table shows the distribution of ratios of Cotality-recorded sale prices to ground truth prices for matched discrepant sales in New York City in 2018. A ``discrepant sale" is any sale where the absolute difference between the broker and ground-truth sale price exceeds 5\% of the ground-truth price.}
\label{tab:cotality_ratio_counts_nyc}
\end{table}

\begin{table*}
\centering
\small
\begin{tabular}{llll}
\toprule
 & Ground Truth & ATTOM & Cotality \\
\midrule
N & 67,343 & 45,504 & 54,278 \\
Mean Sale Price & \$325,492 & \$220,994 & \$254,483 \\
Sale Price P1 & \$16,500 & \$14,000 & \$45,000 \\
Sale Price P25 & \$127,000 & \$100,000 & \$130,000 \\
Sale Price P50 & \$210,000 & \$185,500 & \$210,000 \\
Sale Price P75 & \$320,000 & \$275,000 & \$300,000 \\
Sale Price P99 & \$1,863,147 & \$905,000 & \$1,043,765 \\
Mean Assessed Value & \$197,917 & \$157,063 & \$184,846 \\
Assessed Value P1 & \$14,400 & \$0 & \$17,800 \\
Assessed Value P25 & \$84,200 & \$67,300 & \$88,500 \\
Assessed Value P50 & \$145,800 & \$125,600 & \$144,100 \\
Assessed Value P75 & \$232,000 & \$199,700 & \$216,600 \\
Assessed Value P99 & \$967,990 & \$730,697 & \$822,789 \\
Mean Ratio & 0.88 & 0.91 & 0.79 \\
Ratio P1 & 0.04 & 0.00 & 0.09 \\
Ratio P25 & 0.63 & 0.62 & 0.60 \\
Ratio P50 & 0.80 & 0.81 & 0.76 \\
Ratio P75 & 0.96 & 1.01 & 0.93 \\
Ratio P99 & 3.06 & 4.03 & 1.98 \\
\bottomrule
\end{tabular}
\caption{\textbf{Comparison of Sale Prices, Assessed Values, and Assessment Ratios Between Ground-Truth and Broker-Provided Data for Philadelphia, 2018--2021.} The table displays summary statistics (counts, means, and percentile values) for each of the Philadelphia (ground truth), ATTOM, and Cotality datasets for arms-length, single-family home transactions from 2018 to 2021. ``Ratio" indicates the ratio between the recorded assessed value and property sale price.}
\label{tab:summ_stats_philly}
\end{table*}

\begin{table*}
\centering
\small
\begin{tabular}{lccc}
\toprule
 & \textbf{2018} & \textbf{2019} & \textbf{2020} \\
\midrule
N matched & 2,233 & 11,975 & 12,195 \\
Matched as \% of ground-truth sales & 16.04\% & 74.50\% & 75.23\% \\
\addlinespace
Share with sale price error $>$5\% & 0.76\% & 0.68\% & 0.80\% \\
\quad Conditional mean error (\%) & 211.84\% & 316.49\% & 265.28\% \\
  & (43.20\% - 380.47\%) & (62.07\% - 570.91\%) & (59.17\% - 471.39\%) \\
\quad Conditional mean error (\$) & \$369,814 & \$353,485 & \$1,415,608 \\
  & (\$65,617 - \$674,010) & (\$90,848 - \$616,123) & (\$1,076,582 - \$1,754,634) \\
\addlinespace
Share with assessed value error $>$5\% & 72.28\% & 49.29\% & 1.71\% \\
\quad Conditional mean error (\%) & 60510.57\% & 10128.35\% & 53.02\% \\
  & (-47898.93\% - 168920.08\%) & (-7874.41\% - 28131.11\%) & (40.59\% - 65.45\%)  \\
\quad Conditional mean error (\$) & \$33,307 & \$24,150 & \$189,033 \\
  & (\$30,769 - \$35,846) & (\$22,707 - \$25,594) & (\$163,778 - \$214,287) \\
\bottomrule
\end{tabular}
\caption{\textbf{Discrepancies between ATTOM and Ground Truth Sale Prices and Assessments in Philadelphia, 2018--2020.} The table displays summary statistics of the discrepancies between ATTOM and ground-truth transaction data for sales of single-family homes in Philadelphia from 2018 to 2020. “Error” is calculated as the absolute difference between the data broker and ground truth transaction price or assessed value as a fraction of the ground truth price or value. Conditional means are computed using only matched transactions with error $>$5\%. Records were matched using property address, latitude, longitude, and sale date. 95\% confidence intervals are displayed in parentheses.}
\label{tab:attom_discrepancies_by_year_philly}
\end{table*}

\begin{table}
\centering
\small
\begin{tabular}{lrr}
\toprule
Ratio & Count & Share of Discrepancies (\%) \\
\midrule
10.000 & 18 & 9.18 \\
0.100 & 17 & 8.67 \\
1.122 & 3 & 1.53 \\
1.133 & 3 & 1.53 \\
0.500 & 3 & 1.53 \\
0.123 & 2 & 1.02 \\
0.947 & 2 & 1.02 \\
1.250 & 2 & 1.02 \\
0.010 & 2 & 1.02 \\
1.406 & 1 & 0.51 \\
Other (random) & 123 & 62.76 \\
Other (simple ratio) & 20 & 10.20 \\
Total & 196 & 100.00\% \\
\bottomrule
\end{tabular}
\caption{\textbf{Distribution of Ratios of ATTOM to Ground Truth Sale Prices for Philadelphia Discrepant Sales, 2018--2020.} The table shows the distribution of ratios of Cotality-recorded sale prices to ground truth prices for matched discrepant sales in Philadelphia from 2018 to 2020. A ``discrepant sale" is any sale where the absolute difference between the broker and ground-truth sale price exceeds 5\% of the ground-truth price.}
\label{tab:attom_ratio_counts_philly}
\end{table}

\begin{table*}
\centering
\small
\begin{tabular}{lccc}
\toprule
 & \textbf{2019} & \textbf{2020} & \textbf{2021} \\
\midrule
N matched & 10,903 & 11,505 & 15,256 \\
Matched as \% of ground-truth sales & 67.83\% & 70.97\% & 72.17\% \\
\addlinespace
Share with sale price error $>$5\% & 0.50\% & 0.58\% & 0.64\% \\
\quad Conditional mean error (\%) & 383.48\% & 668.63\% & 1128.27\% \\
   & (11.32\% - 755.63\%) & (168.29\% - 1168.98\%) & (524.43\% - 1732.11\%) \\
\quad Conditional mean error (\$) & \$118,958 & \$434,888 & \$706,586 \\
  & (\$68,177 - \$169,738) & (\$93,020 - \$776,756) & (\$267,201 - \$1,145,970) \\
\addlinespace
Share with assessed value error $>$5\% & 48.10\% & 2.21\% & 1.72\% \\
\quad Conditional mean error (\%) & 10290.76\% & 59.99\% & 55.13\% \\
  & (-9855.46\% - 30436.99\%) & (49.65\% - 70.32\%) & (50.58\% - 59.68\%) \\
\quad Conditional mean error (\$) & \$29,981 & \$235,259 & \$272,150 \\
  & (\$27,670 - \$32,292) & (\$207,986 - \$262,531) & (\$237,504 - \$306,796) \\
\bottomrule
\end{tabular}
\caption{\textbf{Discrepancies between Cotality and Ground Truth Sale Prices and Assessments in Philadelphia, 2019--2021.} The table displays summary statistics of the discrepancies between Cotality and ground-truth transaction data for sales of single-family homes in Philadelphia from 2019 to 2021. “Error” is calculated as the absolute difference between the data broker and ground truth transaction price or assessed value as a fraction of the ground truth price or value. Conditional means are computed using only matched transactions with error $>$5\%. Records were matched using property address, latitude, longitude, and sale date. 95\% confidence intervals are displayed in parentheses.}
\label{tab:cotality_discrepancies_by_year_philly}
\end{table*}

\begin{table*}
\centering
\small
\begin{tabular}{lrr}
\toprule
Ratio & Count & Share of Discrepancies (\%) \\
\midrule
0.100 & 15 & 6.82 \\
0.010 & 11 & 5.00 \\
0.500 & 8 & 3.64 \\
1.133 & 4 & 1.82 \\
0.833 & 4 & 1.82 \\
0.038 & 3 & 1.36 \\
0.305 & 3 & 1.36 \\
1.095 & 3 & 1.36 \\
10.000 & 2 & 0.91 \\
0.916 & 2 & 0.91 \\
Other (random) & 147 & 66.82 \\
Other (simple ratio) & 18 & 8.18 \\
Total & 220 & 100.00\% \\
\bottomrule
\end{tabular}
\caption{\textbf{Distribution of Ratios of Cotality to Ground Truth Sale Prices for Philadelphia Discrepant Sales, 2019--2021.} The table shows the distribution of ratios of Cotality-recorded sale prices to ground truth prices for matched discrepant sales in Philadelphia from 2019 to 2021. A ``discrepant sale" is any sale where the absolute difference between the broker and ground-truth sale price exceeds 5\% of the ground-truth price.}
\label{tab:cotality_ratio_counts_philly}
\end{table*}
\section{Cook County Data Processing} \label{app:data_proc}

Data from Cook County were obtained from the Cook County Open Data Catalog\footnote{https://datacatalog.cookcountyil.gov/stories/s/9bqn-cfsv}. 

The following filters were applied to Parcel Sale data:

\begin{enumerate}
    \item Drop sales occurring before 2018 and after 2021 (`year' $\in [2018, 2021]$)
    \item Drop sales of anything other than single-family homes according to Cook County's residential assessment class designations\footnote{Cook County's residential property class designations are available at: \url{https://prodassets.cookcountyassessoril.gov/s3fs-public/form_documents/classcode.pdf}.}. (`class' $\in [202, 203, 204, 205, 206, 207, 208, 209, 210, 234, 278, 295])$
    \item Subset to deed types corresponding to arms-length transactions (`sale\_filter\_deed\_type' $=$ False)
    \item Drop multi-parcel sales (`is\_multisale' = False)
    \item Drop sales with transaction prices less than \$10K (`sale\_filter\_less\_than\_10k' = False)
    \item Drop duplicate sales (`sale\_filter\_same\_sale\_within\_365' = False, and any combination of `pin' and `sale\_year' occurring more than once in a year) 
    \item Drop any sale missing address, sale, date, or sale amount (`sale\_date', `sale\_price', and `loc\_property\_address' all non null and `loc\_property\_address' $!=$ `0 UNKNOWN')
\end{enumerate}

Sales are joined to assessment records using Cook County property IDs (`pin') and matches between sale year (`year') and assessed year (`tax\_year').
\newpage
\section{Procedure for Matching Brokered and Ground-Truth Data}\label{app:match_procedure}

We first extract street numbers and street names from addresses in both datasets using regular expressions. We then perform an initial match on street number, street name, and sale year.

This initial match includes invalid matches, which we filter out using a fuzzy match procedure. For each match, we compute string similarity scores based on Gestalt pattern matching between broker and ground-truth addresses. We also generate 0-1 indicators for whether the month of sale and latitude/longitude coordinates match across data sources. Among duplicate matches between sales across databases, we keep only those matches that are most similar according to these three criteria. Finally, we drop any match where the address similarity measure falls below 0.8, or the difference in the month of sale listed in each database exceeds 3 months. 

\newpage
\subsection{Robustness of Match Procedure to Alternative Fuzzy Match Thresholds}\label{appendix:match-robustness}

Our results may be sensitive to our choice of address and sale month similarity thresholds. In particular, if our matching thresholds are too generous, our matched sample may include invalid matches which lead us to overstate the extent of sale price and assessed value misreporting. On the other hand, while stricter cutoffs reduce the chance that unrelated transactions match, they may also lead us to drop valid matches and overstate the extent of coverage error in the data. 

To assess the sensitivity of our results to our fuzzy match design, we replicate our match using slightly stricter cutoffs, accepting any match with address similarity greater than 0.9 and with a difference of no more than 2 months between the sale dates listed across data sources.

\subsubsection{ATTOM}

After tightening our matching constraints, we lose 6,546 matches between ground-truth transactions and ATTOM data, decreasing our overall match rate from 87.8\% to 83.6\%. Of these 6,546 transactions, 109 have significant discrepancies between broker-reported and ground-truth sale price, indicating that these matches are predominantly valid. Moreover, the error rate among these transactions is comparable to the overall error rate in our initial matched set (1.7\%). Finally, as shown in Table \ref{tab:attom-mismatch-accounting-thresh}, the share of unmatched ground truth transactions accounted for by our match thresholds jumps from 4\% to 28\% after adjusting the thresholds, indicating that these new thresholds largely impact valid matches and cause us to overstate coverage error in the data.
\begin{table*}[ht]
\centering
\small
\adjustbox{max width=\textwidth}{%
\begin{tabular}{lcccc}
\toprule
 & \textbf{2018} & \textbf{2019} & \textbf{2020} & \textbf{2021} \\
\midrule
N matched & 30,944 & 30,417 & 32,773 & 33,747 \\
Matched as \% of ground-truth sales & 81.56\% & 82.77\% & 83.44\% & 86.36\% \\
\addlinespace
Share with sale price error $>$5\% & 1.25\% & 0.73\% & 3.69\% & 0.15\% \\
\quad Conditional mean error (\%) & 218.87\% & 79.67\% & 64.86\% & 74.56\% \\
  & (107.37\% - 330.38\%) & (64.90\% - 94.44\%) & (59.70\% - 70.03\%) & (37.43\% - 111.70\%) \\
\quad Conditional mean error (\$) & \$315,982 & \$170,913 & \$203,148 & \$164,617 \\
  & (\$230,746 - \$401,218) & (\$141,018 - \$200,808) & (\$190,250 - \$216,046) & (\$124,647 - \$204,587) \\
\addlinespace
Share with assessed value error $>$5\% & 0.01\% & 0.01\% & 0.06\% & 0.97\% \\
\quad Conditional mean error (\%) & 29.50\% & 49.93\% & 55.45\% & 13.42\% \\
  & (-3.53\% - 62.52\%) & (-8.04\% - 107.91\%) & (-16.74\% - 127.63\%) & (12.59\% - 14.26\%) \\
\quad Conditional mean error (\$) & \$12,284 & \$23,214 & \$34,597 & \$4,222 \\
  & (\$-6,789 - \$31,357) & (\$-22,997 - \$69,425) & (\$-21,129 - \$90,323) & (\$3,947 - \$4,498) \\
\bottomrule
\end{tabular}
}%
\caption{\textbf{Discrepancies between ATTOM and Ground Truth Sale Prices and Assessments in Cook County, 2018-2021 (Alternative Matching Thresholds)} The table displays summary statistics of the discrepancies between ATTOM and ground-truth transaction data for sales of single-family homes in Cook County from 2018–2021. “Error” is calculated as the absolute difference between the data broker and ground truth transaction price or assessed value as a fraction of the ground truth price or value. Conditional means are computed using only matched transactions with error $>$5\%. Records were matched using property address, latitude, longitude, and sale date. 95\% confidence intervals are displayed in parentheses.}
\label{tab:attom_discrepancies_by_year_thresh}

\end{table*}
\begin{comment}
\vspace{4pt} % small vertical space
\begin{minipage}{\textwidth}
\scriptsize
\textit{Notes}: The table displays summary statistics of the discrepancies between ATTOM and ground-truth transaction data for sales of single-family homes in Cook County from 2018–2021. “Error” is calculated as the absolute difference between the data broker and ground truth transaction price or assessed value as a fraction of the ground truth price or value. Conditional means are computed using only matched transactions with error $>$5\%. Records were matched using property address, latitude, longitude, and sale date. 95\% confidence intervals are displayed in parentheses.
\end{minipage}
\end{comment}

\setlength\tabcolsep{0.04\linewidth}
\begin{table}[]
\small
\centering
\begin{tabular}{lrr}
\toprule
Ratio & Count & Share of Discrepancies (\%) \\
\midrule
0.667 & 954 & 50.96 \\
2.000 & 147 & 7.85 \\
0.333 & 111 & 5.93 \\
0.666 & 90 & 4.81 \\
0.833 & 28 & 1.50 \\
0.665 & 15 & 0.80 \\
0.250 & 14 & 0.75 \\
10.000 & 12 & 0.64 \\
1.999 & 11 & 0.59 \\
1.133 & 10 & 0.53 \\
Other (random) & 400 & 21.37 \\
Other (simple ratio) & 80 & 4.27 \\
Total & 1,872 & 100.00\% \\
\bottomrule
\end{tabular}
  
\caption{\textbf{Distribution of Ratios of ATTOM to Ground Truth Sale Prices for Cook County Discrepant Sales, 2018-2021 (Alternative Matching Thresholds)} The table shows the distribution of ratios of ATTOM-recorded sale prices to ground truth prices for matched discrepant sales in Cook County covering 2018-2021. A ``discrepant sale" is any sale where the absolute difference between the broker and ground-truth sale price exceeds 5\% of the ground-truth price.}
\begin{comment}
\vspace{4pt} % small vertical space
\begin{minipage}{\textwidth}
\scriptsize
      \scriptsize
      \textit{Notes}: The table shows the distribution of ratios of ATTOM-recorded sale prices to ground truth prices for matched discrepant sales in Cook County covering 2018-2021. A ``discrepant sale" is any sale where the absolute difference between the broker and ground-truth sale price exceeds 5\% of the ground-truth price. 
    \end{minipage}
    \label{tab:attom_ratio_counts}
\end{comment}
\end{table}

\begin{table}[ht]
\centering
\small
\adjustbox{max width=\textwidth}{%

\begin{tabular}{lr}
\textbf{N ground truth sales} & 153,044 \\ 
\textbf{N sales not matched to ATTOM} & 25,163 \\ 
\textbf{Reason for match failure:} &  \\ 
Record missing & 27.2\% \\ 
Matches duplicate sale & 4.9\% \\ 
Broker filter misreported & 39.1\% \\ 
\quad Arms-length sale & 30.2\% \\ 
\quad Not multiparcel & 2.1\% \\ 
\quad Sale price $>$ \$10k & 12.0\% \\ 
\quad Single-family home & 12.0\% \\ 
Does not meet fuzzy match criteria & 28.2\% \\ 
Other & 0.5\% \\ 
\hline 
\textbf{Total} & 100.0\% 
\end{tabular}
}%
\caption{\textbf{Sources of coverage error: ATTOM (Alternative Matching Thresholds)} The table displays the sources of coverage error between ground truth and attom data for arms-length sales of single family homes in Cook County from 2018-2021.}
\label{tab:attom-mismatch-accounting-thresh}
\begin{comment}
\vspace{4pt} % small vertical space
\begin{minipage}{\textwidth}
\scriptsize
\textit{Notes}: The table displays the sources of coverage error between ground truth and attom data for arms-length sales of single family homes in Cook County from 2018-2021. 
\end{minipage}
\end{comment}
\end{table}

\subsubsection{Cotality}

Tightening our match conditions leads to 4,385 fewer matches between Cotality and ground-truth data. Among these transactions, 94 have significant errors in reported sale price, for an overall error rate of roughly 2.1\%. As with ATTOM, this error rate is comparable to that of the initial matched set (1.5\%), indicating that stricter match conditions lead us to drop predominantly valid matches. Moreover, as shown in Table \ref{tab:cotality-mismatch-accounting-thresh}, the share of unmatched transactions accounted for by our match conditions jumps to 17.8\%, indicating that stricter match conditions penalize valid matches and lead us to overstate the extent of coverage error.\looseness=-1

\begin{table*}[ht]
\centering
\small
\adjustbox{max width=\textwidth}{%
\begin{tabular}{lcccc}
\toprule
 & \textbf{2018} & \textbf{2019} & \textbf{2020} & \textbf{2021} \\
\midrule
N matched & 29,609 & 30,067 & 32,915 & 32,630 \\
Matched as \% of ground-truth sales & 78.04\% & 81.82\% & 83.80\% & 83.50\% \\
\addlinespace
Share with sale price error $>$5\% & 1.03\% & 0.59\% & 3.67\% & 0.41\% \\
\quad Conditional mean error (\%) & 93.67\% & 90.77\% & 68.57\% & 138.12\% \\
  & (77.07\% - 110.26\%) & (72.38\% - 109.17\%) & (62.63\% - 74.51\%) & (95.41\% - 180.82\%) \\
\quad Conditional mean error (\$) & \$213,823 & \$162,251 & \$206,753 & \$453,650 \\
  & (\$165,086 - \$262,560) & (\$131,469 - \$193,034) & (\$191,763 - \$221,742) & (\$299,766 - \$607,534) \\
\addlinespace
Share with assessed value error $>$5\% & 0.01\% & 0.00\% & 0.06\% & 0.98\% \\
\quad Conditional mean error (\%) & 37.49\% & 0.00\% & 58.69\% & 15.41\% \\
  & (-18.19\% - 93.18\%) & (0.00\% - 0.00\%) & (-17.31\% - 134.69\%) & (13.11\% - 17.71\%) \\
\quad Conditional mean error (\$) & \$22,351 & \$0 & \$33,800 & \$4,792 \\
  & (\$-14,816 - \$59,519) & (\$0 - \$0) & (\$-25,109 - \$92,710) & (\$4,192 - \$5,392) \\
\bottomrule
\end{tabular}
}%
\caption{\textbf{Discrepancies between Cotality and Ground Truth Sale Prices and Assessments in Cook County, 2018-2021 (Alternative Matching Thresholds)} The table displays summary statistics of the discrepancies between Cotality and ground-truth transaction data for sales of single-family homes in Cook County from 2018–2021. “Error” is calculated as the absolute difference between the data broker and ground truth transaction price or assessed value as a fraction of the ground truth price or value. Conditional means are computed using only matched transactions with error $>$5\%. Records were matched using property address, latitude, longitude, and sale date. 95\% confidence intervals are displayed in parentheses.}
\label{tab:cotality_discrepancies_by_year_thresh}

\end{table*}
\begin{comment}
\vspace{4pt} % small vertical space
\begin{minipage}{\textwidth}
\scriptsize
\textit{Notes}: The table displays summary statistics of the discrepancies between ATTOM and ground-truth transaction data for sales of single-family homes in Cook County from 2018–2021. “Error” is calculated as the absolute difference between the data broker and ground truth transaction price or assessed value as a fraction of the ground truth price or value. Conditional means are computed using only matched transactions with error $>$5\%. Records were matched using property address, latitude, longitude, and sale date. 95\% confidence intervals are displayed in parentheses.
\end{minipage}
\end{comment}

\setlength\tabcolsep{0.04\linewidth}
\begin{table}[]
\small
\centering
\begin{tabular}{lrr}
\toprule
Ratio & Count & Share of Discrepancies (\%) \\
\midrule
0.667 & 1,038 & 56.88 \\
0.333 & 134 & 7.34 \\
0.666 & 96 & 5.26 \\
2.000 & 44 & 2.41 \\
0.833 & 28 & 1.53 \\
0.250 & 17 & 0.93 \\
0.095 & 14 & 0.77 \\
0.948 & 11 & 0.60 \\
0.665 & 10 & 0.55 \\
1.133 & 9 & 0.49 \\
Other (random) & 336 & 18.41 \\
Other (simple ratio) & 88 & 4.82 \\
Total & 1,825 & 100.00\% \\
\bottomrule
\end{tabular}
  
\caption{\textbf{Distribution of Ratios of cotality to Ground Truth Sale Prices for Cook County Discrepant Sales, 2018-2021 (Alternative Matching Thresholds)} The table shows the distribution of ratios of cotality-recorded sale prices to ground truth prices for matched discrepant sales in Cook County covering 2018-2021. A ``discrepant sale" is any sale where the absolute difference between the broker and ground-truth sale price exceeds 5\% of the ground-truth price.}
\begin{comment}
\vspace{4pt} % small vertical space
\begin{minipage}{\textwidth}
\scriptsize
      \scriptsize
      \textit{Notes}: The table shows the distribution of ratios of cotality-recorded sale prices to ground truth prices for matched discrepant sales in Cook County covering 2018-2021. A ``discrepant sale" is any sale where the absolute difference between the broker and ground-truth sale price exceeds 5\% of the ground-truth price. 
    \end{minipage}
    \label{tab:cotality_ratio_counts}
\end{comment}
\end{table}

\begin{table}[ht]
\centering
\small
\adjustbox{max width=\textwidth}{%

\begin{tabular}{lr}
\textbf{N ground truth sales} & 153,044 \\ 
\textbf{N unmatched sales} & 27,823 \\ 
\textbf{Share of unmatched:} &  \\ 
Record missing & 47.2\% \\ 
Matches duplicate sale & 7.1\% \\ 
Broker filter misreported & 27.3\% \\ 
\quad Arms-length sale & 24.3\% \\ 
\quad Not multiparcel & 2.6\% \\ 
\quad Sale price $>$ \$10k & 7.8\% \\ 
\quad Single-family home & 7.8\% \\ 
Does not meet fuzzy match criteria & 17.8\% \\ 
Other & 0.5\% \\ 
\hline 
\textbf{Total} & 100.0\% 
\end{tabular}
}%
\caption{\textbf{Sources of coverage error: Cotality (Alternative Matching Thresholds)} The table displays the sources of coverage error between ground truth and Cotality data for arms-length sales of single family homes in Cook County from 2018-2021.}
\label{tab:cotality-mismatch-accounting-thresh}
\begin{comment}
\vspace{4pt} % small vertical space
\begin{minipage}{\textwidth}
\scriptsize
\textit{Notes}: The table displays the sources of coverage error between ground truth and Cotality data for arms-length sales of single family homes in Cook County from 2018-2021. 
\end{minipage}
\end{comment}
\end{table}
\newpage
\section{Regressivity Metrics}\label{appendix:regressivity-metrics}

Below, we provide a brief overview of the intuition and formulas for each of the regressivity metrics in the paper.

\subsection{Price-Related Differential (PRD)}

The Price-Related Differential is the ratio between the mean assessment ratio and the mean assessment ratio weighted by sale price. It is given by:

\begin{equation}
    PRD = \frac{\bar{R}}{\sum_{i}(\frac{{A_i}}{{S_i}})(\frac{S_i}{\sum_j{S_j}})} = \frac{\bar{R}}{\bar{A}/\bar{S}}
\end{equation}

Where $\bar{R}$ is the mean ratio of assessed values to sale prices, $\bar{A}$ is the mean assessed value and $\bar{S}$ is the mean sale price.

PRD values greater than 1 indicate that high-priced homes are under-assessed relative to low-priced homes. To see why, consider that high-value homes receive a higher weight in the denominator of the PRD. Upweighting smaller ratios pushes the value of the denominator below that of the numerator, generating a ratio greater than 1. Conversely, PRD values less than 1 indicate that low-priced homes are under-assessed relative to high priced homes. Standards established by the International Association of Assessing Officers (IAAO) stipulate that acceptable values for the PRD lie between 0.98 and 1.03 \citep{iaaoStandard2017}.

\subsection{Suits Index}

The Suits Index, first developed by \cite{suits1977measurement} as a measure of tax progressivity, is conceptually similar to the Gini Ratio in that its derivation relies on Lorenz curves. In contrast to the Gini ratio, the Suits Index is derived from a single curve and is calculated as follows:

\begin{enumerate}
    \item Sort homes in ascending order by sale price
    \item Calculate the accumulated percentage of sale prices and assessed values across this sorted list; these percentages will range from 0 to 100.
    \item Plot accumulated assessed values on the y-axis against accumulated sale price on the x-axis.
    \item The suits index is given by:
    \begin{equation}
        S = \frac{(K-L)}{K}
    \end{equation}
    Where K is the area under the 45-degree line, and L is the area under the accumulated value curve. Because the X and Y axes range from 0 to 100, K is equal to $0.5*100*100 = 5,000$. The suits index therefore becomes $S = 1-\frac{1}{5000}*L$.
\end{enumerate}

We calculate the area $L$ using a discrete approximation described in \cite{mcmillenMeasuresVerticalInequality2023}. A Suits Index of 0 indicates that assessed values are perfectly proportional to sale prices ($K=L$). If $L>K$, then the lowest e.g. 10\% of sale value accounts for more than 10\% of assessed value, and assessments are regressive. In this instance, the Suits Index would be negative. Conversely, a positive Suits Index indicates assessments are progressive.

\subsection{Regression Coefficient}

The regression-based measure presented here represents the coefficient $\beta_1$ from the following regression of log sale prices on log sales ratios:

\begin{equation}
    log(A/S) = \beta_0 + \beta_1 log(S) + \varepsilon
\end{equation}

Where $A$ is assessed value and $S$ is sale price. If assessments are regressive, then log sales ratios will generally decrease with log sale prices, and $\beta_1$ will be negative.

As \cite{mcmillenMeasuresVerticalInequality2023} point out, regression-based measures of assessment inequality are generally biased downwards, towards finding regressivity. This bias is attributable in part to measurement error, overlap between the data used to train and evaluate assessment models, and the functional form of assessment models. That being said, as the simulations in \cite{mcmillenMeasuresVerticalInequality2023} demonstrate, regression-based measures can effectively capture changes in simulated regressivity even if the levels are biased. In this application, we are  interested in both changes as well as absolute levels of regressivity as we move from ground truth to broker-provided data, and for this application the measure, though biased, remains informative. We present alternative measures for robusttness.


\end{document}